\documentclass[twoside]{article}

\usepackage{amsmath}
\usepackage{amssymb}
\usepackage{amscd}
\usepackage{epsfig}
\begin{document}

\newtheorem{defin}{Definition}
\newtheorem{definition}{Definition}
\newtheorem{corol}{Corollaire}
\newtheorem{exemple}{Exemple}
\newtheorem{exercice}{Exercice}
\newtheorem{lemme}{Lemme}
\newtheorem{propos}{Proposition}
\newtheorem{theor}{\large\bf Theorem}
\newtheorem{theorem}{Theorem}[section]
 \newtheorem{lemma}[theorem]{Lemma}
 \newtheorem{prop}[theorem]{Proposition}
 \newtheorem{cor}[theorem]{Corollary}
 \newtheorem{fact}[theorem]{Fact}
 \newtheorem{conj}[theorem]{Conjecture}
 \newtheorem{quest}[theorem]{Question}
 \newtheorem{defi}[theorem]{Definition}
 \newtheorem{example}[theorem]{Example}
 \newtheorem{remark}[theorem]{Remark}
 \newtheorem{exercise}[theorem]{Exercise}

\def\k{{\bf k}}
\def\q{{\bf q}}
\def\x{{\bf x}}

\newcommand{\qed}{{\hfill $\Box$}}
\newcommand{\vep}{\varepsilon}
\newcommand{\mb}[1]{\mbox{$#1$}}
\newcommand{\prf}{\noindent{\bf Proof}\ }

% raccourcis pour les lettres "script"
\newcommand\scA{{\mathscr A}}
\newcommand\scB{{\mathscr B}}
\newcommand\scC{{\mathscr C}}
\newcommand\scD{{\mathscr D}}
\newcommand\scE{{\mathscr E}}
\newcommand\scF{{\mathscr F}}
\newcommand\scG{{\mathscr G}}
\newcommand\scH{{\mathscr H}}
\newcommand\scI{{\mathscr I}}
\newcommand\scJ{{\mathscr J}}
\newcommand\scK{{\mathscr K}}
\newcommand\scL{{\mathscr L}}
\newcommand\scM{{\mathscr M}}
\newcommand\scN{{\mathscr N}}
\newcommand\scO{{\mathscr O}}
\newcommand\scP{{\mathscr P}}
\newcommand\scQ{{\mathscr Q}}
\newcommand\scR{{\mathscr R}}
\newcommand\scS{{\mathscr S}}
\newcommand\scT{{\mathscr T}}
\newcommand\scU{{\mathscr U}}
\newcommand\scV{{\mathscr V}}
\newcommand\scW{{\mathscr W}}
\newcommand\scX{{\mathscr X}}
\newcommand\scY{{\mathscr Y}}
\newcommand\scZ{{\mathscr Z}}

% raccourcis pour les lettres calligraphiees
\newcommand\cA{{\mathcal A}}
\newcommand\cB{{\mathcal B}}
\newcommand\cC{{\mathcal C}}
\newcommand\cD{{\mathcal D}}
\newcommand\cE{{\mathcal E}}
\newcommand\cF{{\mathcal F}}
\newcommand\cG{{\mathcal G}}
\newcommand\cH{{\mathcal H}}
\newcommand\cI{{\mathcal I}}
\newcommand\cJ{{\mathcal J}}
\newcommand\cK{{\mathcal K}}
\newcommand\cL{{\mathcal L}}
\newcommand\cM{{\mathcal M}}
\newcommand\cN{{\mathcal N}}
\newcommand\cO{{\mathcal O}}
\newcommand\cP{{\mathcal P}}
\newcommand\cQ{{\mathcal Q}}
\newcommand\cR{{\mathcal R}}
\newcommand\cS{{\mathcal S}}
\newcommand\cT{{\mathcal T}}
\newcommand\cU{{\mathcal U}}
\newcommand\cV{{\mathcal V}}
\newcommand\cW{{\mathcal W}}
\newcommand\cX{{\mathcal X}}
\newcommand\cY{{\mathcal Y}}
\newcommand\cZ{{\mathcal Z}}

% raccourcis pour les lettres "gothiques"
\newcommand\kb{{\mathfrak b}}
\newcommand\kc{{\mathfrak c}}
\newcommand\kf{{\mathfrak f}}
\newcommand\kg{{\mathfrak g}}
\newcommand\kh{{\mathfrak h}}
\newcommand\ki{{\mathfrak i}}
\newcommand\km{{\mathfrak m}}
\newcommand\ko{{\mathfrak{o}}}
\newcommand\ks{{\mathfrak{s}}}
\newcommand\kt{{\mathfrak{t}}}
\newcommand\ku{{\mathfrak{u}}}
\newcommand\kv{{\mathfrak{v}}}
\newcommand\kw{{\mathfrak w}}

\newcommand\kA{{\mathfrak A}} 
\newcommand\kB{{\mathfrak B}} 
\newcommand\kC{{\mathfrak C}} 
\newcommand\kF{{\mathfrak F}}
\newcommand\kG{{\mathfrak G}}
\newcommand\kH{{\mathfrak H}}
\newcommand\kL{{\mathfrak L}} 
\newcommand\kP{{\mathfrak P}} 
\newcommand\kS{{\mathfrak S}} 
\newcommand\kT{{\mathfrak T}}
\newcommand\kU{{\mathfrak U}}
\newcommand\kV{{\mathfrak V}}
\newcommand\kW{{\mathfrak W}}
\newcommand\kX{{\mathfrak X}}

\newcommand\kgl{{\mathfrak{gl}}} 
\newcommand\ksl{{\mathfrak{sl}}}
\newcommand\kso{{\mathfrak{so}}}
\newcommand\ksu{{\mathfrak{su}}}

\newcommand\Ug{{\cU(\kg)}}

\newcommand{\eqdef} {\stackrel{\rm def}{=}}

\newcommand{\Br}  {\overline}
\newcommand{\R}{{\mathbb R}}
\newcommand{\N}{{\mathbb N}}
\newcommand{\Q}{{\mathbb Q}}
\newcommand{\ZZ}{{\mathbb Z}}
\newcommand{\C}{{\mathbb C}}
\newcommand{\be}{\begin{equation}}
\newcommand{\ee}{\end{equation}}
\newcommand{\bea}{\begin{eqnarray}}
\newcommand{\eea}{\end{eqnarray}}
\newcommand{\bqa}{\begin{eqnarray}}
\newcommand{\eqa}{\end{eqnarray}}
\newcommand{\ba}{\begin{array}}
\newcommand{\ea}{\end{array}}
\newcommand{\p}[1]{{\partial\over \partial{#1}}}
\newcommand{\no}{\nonumber}
\newcommand{\sq}{\sqrt{2}}
\newcommand{\da}{\dagger}
\newcommand{\lp}{\left (}
\newcommand{\rp}{\right )}
\newcommand{\al}{\alpha}
\newcommand{\bt}{\beta}
\newcommand{\ga}{\gamma}
\newcommand{\de}{\delta}
\newcommand{\e}{\epsilon}
\newcommand{\ze}{\zeta}
\newcommand{\te}{\theta}
\newcommand{\io}{\iota}
\newcommand{\ka}{\kappa}
\newcommand{\la}{\lambda}
\newcommand{\ta}{\tau}
\newcommand{\ro}{\rho}
\newcommand{\ps}{\psi}
\newcommand{\ch}{\chi}
\newcommand{\ep}{\epsilon}
\newcommand{\si}{\sigma}
\newcommand{\up}{\upsilon}
\newcommand{\om}{\omega}
\newcommand{\Om}{\Omega}
\newcommand{\Si}{\Sigma}
\newcommand{\La}{\Lambda}
\newcommand{\Lazero}{\Lambda_{0}}
\newcommand{\Lainv}{{\Lambda^{-2}}}
\newcommand{\Lazeroinv}{{\Lambda^{-2}_0}}
\newcommand{\Ga}{\Gamma}
\newcommand{\De}{\Delta}
\newcommand{\Th}{\Theta}
\newcommand{\Up}{\Upsilon}
\newcommand{\bpsi}{\bar{\psi}}

\newcommand{\optbar}{\bar}

\def\FF{\hbox to 8.33887pt{\rm I\hskip-1.8pt F}}
\def\NN{\hbox to 9.3111pt{\rm I\hskip-1.8pt N}}
\def\PP{\hbox to 8.61664pt{\rm I\hskip-1.8pt P}}
\def\QQ{\rlap {\raise 0.4ex \hbox{$\scriptscriptstyle |$}}
{\hskip -4.5pt Q}}
\def\RR{\hbox to 9.1722pt{\rm I\hskip-1.8pt R}}
\def\ZZ{\hbox to 8.2222pt{\rm Z\hskip-4pt \rm Z}}
\def\ZZZ{Z\!\!\!Z}

\numberwithin{equation}{section}

\title
{Introduction to the Renormalization Group\\ with Applications to
Non-Relativistic\\ 
Quantum Electron Gases}
\author{CIME Lectures, Cetraro\\
\\
Vincent Rivasseau}
\maketitle

\begin{abstract}
We review the rigorous work on many Fermions models
which lead to the first constructions of interacting
Fermi liquids in two dimensions,
and allowed to prove that there are different
scaling regimes in two dimensions, depending
on the shape of the Fermi surface. We also review progress on the three
dimensional case. 

We start with a pedagogical introduction on quantum field theory 
and perturbative renormalization. Emphasis is then put on using 
renormalization around the Fermi surface in a constructive way, in which 
all orders of perturbation theory are summed rigorously. 

%We end up with a brief overview on two other topics of current interest to us, namely non commutative field theory and group field theory, that could benefit from interaction with specialists of renormalization theory.
 
\end{abstract}

\section{Introduction to QFT and Renormalization}

\subsection{Gaussian Measures}

A finite dimensional centered normalized Gaussian measure $d\mu_C$
is defined through its covariance. 
Consider a finite dimensional space ${\mathbb R}^N$ and a symmetric
positive definite $N$ by $N$ matrix $A$. The inverse of the matrix $A$
is also a definite positive symmetric  $N$ by $N$ matrix $C= A^{-1}$ called
the covariance associated to $A$.
The corresponding
centered normalized Gaussian measure is
\bea  d\mu_C = (2\pi)^{-N/2}  \sqrt{\det A }  \;\;  e^{- \frac{1}{2} ^tXAX } d^N X ,  \label{gauss1}
\eea
so that $\int  d\mu_C = 1$.

To understand $d\mu_C$ it is better to know $C$ than $A$
since the moments or correlation functions of a Gaussian measure can
be expressed simply as sums of monomials in $C$. In fact formula \eqref{gauss1} perfectly
makes sense if $C$ is non invertible, and even for $C=0$; but the corresponding measure
has no density with respect to the Lebesgue measure in this case (for $C=0$
$d\mu_C$ is just Dirac's $\delta$ function at the origin).
The reader familiar with eg ordinary linear PDE's knows that the essential
point is to invert the matrix or the operator, hence to know the ``Green's function".
But quadratic forms have linear equations as their variational solutions, so both problems are linked.

Sine any function can be approximated by polynomials,
probability measures are characterized by their moments, that is by the integrals
they return for each polynomial of the integration variables.

The corresponding theorem which computes the moments of a Gaussian
measure in terms of the covariance is fundamental in QFT and known
there under the name of Wick's theorem. It expresses the result as the
sum over all possible pairings or the variables of a
product of covariances between the paired variables:
\bea \label{wick}
\int X_{i_1} ...  X_{i_n} d \mu_C = \sum_G \prod_{\ell \in G} C_{i_{b(\ell)},i_{e(\ell)}} ,
\eea
where $G$ runs over all Wick contractions or pairings
of the labels $1, ..., n$. Each  pair $\ell$ is pictured as a line joining two labels 
$b(\ell)$ and $e(\ell)$ (which we call arbitrarily the ``beginning" and ``end" of the line).

The theory of Gaussian measures and Wick's theorem 
extends to infinite dimensional spaces, in which the covariance
$C$ may become a positive kernel $C(x,y)$
in a distribution space. We recall that such a kernel is an
operator acting on functions through $C.f = \int  C(x,y) f(y)\, dy$.
The identity operator is represented by the 
Dirac kernel $C(x,y) = \delta (x-y)$.
But if $C$ is positive definite with some regularity
it can be also considered the covariance of a truly well
defined Gaussian measure in some infinite dimensional
space of distributions, through an extension of Bochner's theorem
known as Minlos Theorem \cite{GJ}.

\subsection{Functional integrals}

In QFT, like in grand-canonical statistical mechanics, particle number is 
not conserved. Cross sections in scattering experiments contain the physical information of the 
theory\footnote{Correlation
functions play this fundamental role in statistical mechanics}. 
They are the matrix elements of the diffusion matrix $\cS$. 
Under suitable conditions they are expressed in terms of 
the Green functions $G_{N}$ 
of the theory through so-called ``reduction formulae".

Green functions are time ordered vacuum expectation values
of the field $\phi$, which is operator valued and acts on the Fock space:
\be  G_{N}(z_{1},...,z_{N})  =  <\psi_{0}, 
T[\phi(z_{1})...\phi (z_{N})]\psi_{0}> \,  .
\ee
Here $\psi_{0}$ is the vacuum state and the
$T$-product orders 
$\phi(z_{1})...\phi (z_{N})$ according to increasing times.

Consider a Lagrangian field theory, and split 
the total Lagrangian as the sum of a free plus an
interacting piece, $\cL=  \cL_{0} + \cL_{int}$. 
The Gell-Mann-Low formula expresses
the Green functions as vacuum expectation values of a similar product
of free fields with an $e^{i\int \cL_{int}}$ insertion:
\be G_{N}(z_{1},...,z_{N})  = \frac{ <\psi_{0}, 
T\biggl[ \phi(z_{1})...\phi (z_{N})e^{i\int dx \cL_{int}(\phi(x))}\biggr]
 \psi_{0}>}     {<\psi_{0}, 
T(e^{i\int dx \cL_{int}(\phi(x))}) \psi_{0}>    } .    
\ee

In the functional integral formalism proposed by Feynman \cite{FH},
the Gell-Mann-Low formula is 
replaced by a functional integral in terms of an (ill-defined) 
``integral over histories"
which is formally the product of Lebesgue measures over all space time. 
The corresponding formula is the Feynman-Kac formula:
\be G_{N}(z_{1},...,z_{N}) =
  = \frac { \int  \prod\limits_{j}\phi(z_{j}) e^{i\int \cL(\phi(x))
  dx} D\phi}  
{\int   e^{i\int \cL(\phi(x)) dx} D\phi }    .   \label{funct}
\ee

The integrand in (\ref{funct}) contains now 
the full Lagrangian $\cL=\cL_{0}+\cL_{int}$ instead of the interacting one. 
This is  interesting to expose symmetries of the theory which may not be separate 
symmetries of the free and interacting Lagrangians, for instance 
gauge symmetries. Perturbation theory
and the Feynman rules can still be derived as explained in the next subsection. 
But (\ref{funct}) is also well adapted to constrained quantization
and to the study of non-perturbative effects. 

For general references on QFT, see \cite{Pesk,ItZu,Ram}.

\subsection{Statistical Mechanics and Thermodynamic Quantities}

There is a deep analogy between the Feynman-Kac formula and the 
formula which expresses correlation functions in classical statistical mechanics. 

The partition function of a statistical mechanics
grand canonical ensemble described by a Hamiltionian $H$ at temperature $T$ and  
chemical potential $\mu$ is
\be
Z_\Lambda = Tr (e^{-\beta (H- \mu N)}) ,
\ee
where $\beta = 1/kT$, and the trace may be either a classical integration in phase-space
or in the quantum case a trace on the relevant Hilbert space (Fock space).

The main problem is to compute the {\it logarithm} of the partition function.
Indeed thermodynamic quantities such as eg the mean energy of the system
\be
< H >_{T, \mu} =\frac{Tr(He^{-\beta (H - \mu N)})}{Z_\Lambda}= - \frac{\partial \log Z_\lambda}{\partial \beta} ,
\ee
the free energy
\be
F = <H - \mu N >_{T, \mu} = - \log Z_\Lambda ,
\ee
the entropy
\be 
S = \frac{1}{\beta^2} \frac{\partial F}{\partial \beta},
\ee
or the heat capacity (at fixed volume)  
\be C_V  =  \biggl(  \frac{\partial < H >_{T,\mu}}{\partial T} \biggr)_\mu  = - \frac{1}{\beta^2}
 \biggl(  \frac{\partial < H >_{T,\mu}}{\partial \beta} \biggr)_\mu
\ee
follow from that computation. 

In fact all the detailed information on the equilibrium states is encoded in the
list of their {\it correlation functions}, which are derivatives of the
logarithm of the partition function with respect to appropriate sources. 
For instance for a lattice Ising model 
the partition function is 
\be  Z_\Lambda =  \sum\limits_{ \{ \sigma_{x} = \pm 1 \} } e^{- L(\sigma) } 
\ee
and the correlation functions are
\be \bigl<   \prod_{i=1}^{n} \sigma_{x_{i}} \bigr>
  = \frac { \sum\limits_{ \{ \sigma_{x} = \pm 1 \} } e^{- L(\sigma)}
 \prod\limits_{i}\sigma_{x_{i}}} 
{\sum\limits_{ \{ \sigma_{x} = \pm 1 \} } e^{- L(\sigma) }}   ,
\label{ising}    \ee
where $x$ labels the discrete sites of the lattice. The sum
is over configurations $\{ \sigma_{x} = \pm 1 \}$ which associate
a ``spin'' with value +1 or -1 to each such site and 
$L(\sigma)$ contains usually nearest neighbor interactions 
and possibly a magnetic field h: 
\be L(\sigma) = \sum_{x, y \ {\rm nearest\  neighbors}} J \sigma_{x} 
\sigma_{y} + \sum_{x} h \sigma_{x} .\ee

By analytically continuing (\ref{funct}) to imaginary time, or 
Euclidean space, it is possible to complete the analogy with (\ref{ising}), 
hence to establish a firm contact between Euclidean QFT and statistical mechanics
\cite{Zinn,ItDr,Par}.

\subsection{Schwinger Functions}
This idea also allows 
to give much better meaning to the  
path integral, at least for a free Bosonic field. Indeed the
free Euclidean measure  can be defined easily as a Gaussian measure, because
in Euclidean space $L_{0}$ is a quadratic form of positive type\footnote{However 
the functional space
that supports this measure is not in general a space of smooth functions,
but rather of distributions. This was already true for functional integrals
such as those of Brownian motion, which are supported by continuous but
not differentiable paths. Therefore ``functional integrals'' in quantum
field theory should more appropriately be 
called ``distributional integrals''.}.

The Green functions continued to Euclidean points are called the Schwinger 
functions of the model, and are given by the Euclidean Feynman-Kac formula:
\be S_{N}(z_{1},...,z_{N}) =
Z^{-1} \int  \prod_{j=1}^{N} \phi(z_{j}) e^{-\int \cL (\phi(x)) dx} 
D\phi   ,
\ee
\be Z= \int   e^{-\int \cL (\phi(x)) dx} 
D\phi  .\ee

The simplest interacting field theory is the theory of a one component scalar 
bosonic field $\phi$ with quartic interaction $\lambda\phi^{4}$ ($\phi^{3}$, which is 
simpler, is unstable).  In ${\R}^{d}$ it is called the $\phi^{4}_{d}$ 
model. For $d=2,3$ this model is 
superrenormalizable and has been built non perturbatively  by constructive 
field theory (see \cite{GJ,Riv}). In these dimensions the model 
is unambiguously related to its perturbation
series \cite{EMS,MS} through Borel summability \cite{Sok}. 
For $d=4$ the model is just renormalizable, and provides the simplest 
pedagogical introduction to perturbative renormalization theory. But 
because of the Landau ghost or triviality problem explained in subsection 
\ref{Landaughost}, the model presumably does not exist as a true interacting
theory at the non perturbative level (see \cite{Riv} for a discussion of this subtle issue).

Formally the Schwinger functions of $\phi^{4}_{d}$ are the moments of the 
measure:
\be  d\nu  = \frac {1}{Z} e^{-\frac{\lambda}{4!}\int \phi^{4} -(m^{2} / 2)
\int \phi^{2} - (a/2) \int (\partial _{\mu } \phi \partial ^{\mu }\phi  )
} D\phi , \label{mesurenu}\ee
where
\begin{itemize}
\item $\lambda$ is the coupling constant, usually assumed positive or complex 
with positive real part; remark the arbitrary but convenient 1/4! factor to take into account
the symmetry of permutation of all fields at a local vertex.

\item $m$ is the mass, which fixes an energy scale for the theory;

\item $a$ is the wave function constant. It can be set to 1 by a rescaling of the field.

\item $Z$ is a normalization factor which makes (\ref{mesurenu}) a probability 
measure;

\item $D\phi$ is a formal (mathematically ill-defined) product 
$\prod\limits_{x \in {\R}^{d} }d\phi(x)$
of Lebesgue measures at every point of ${\R}^{d}$.
\end{itemize}

The Gaussian part of the measure is
\be  d\mu(\phi)  =  {1 \over Z_0} e^{-(m^{2} / 2)
\int \phi^{2} - (a/2) \int (\partial _{\mu } \phi \partial ^{\mu }\phi  )
} D\phi . \label{mesuremu}
\ee
where $Z_0$ is again the normalization factor which makes
(\ref{mesuremu}) a probability  measure.

More precisely if we consider the translation invariant 
propagator $C(x,y) \equiv C(x-y)$ (with slight abuse of notation), whose 
Fourier transform is

\be  C(p) =  \frac{1}{(2\pi)^{d}}  \frac{1}{p^{2}+m^{2}} , \ee
we can use Minlos theorem and the general theory of Gaussian processes  
to define $d\mu(\phi)$ as the centered Gaussian measure on the
Schwartz space of tempered distributions $S'({\R}^{d})$
whose covariance is $C$. A Gaussian measure is uniquely defined 
by its moments, or the integral of polynomials of fields. 
Explicitly this integral is zero for a monomial of odd degree, and for even $n=2p$
it is equal to 
\be   \int  \phi  (x_1 ) ...\phi  (x_n ) d\mu(\phi) = \sum_{\cW} 
\prod_{\ell \in \cW} C(x_{b(\ell)},x_{e(\ell)}) ,
\ee
where the sum runs over all the $2p!! = (2p-1)(2p-3)...5.3.1$ Wick
pairings $\cW$ of the $2p$ arguments into the $p$ disjoint
pairs $\ell= (b(\ell),e(\ell))$. 

Note that since for $d\ge 2$, $C(p)$ is not integrable,
$C(x,y)$ must be understood as a distribution. It is therefore
convenient to also use regularized kernels, for instance
\be     C_{\kappa}(p) =  \frac{1}{(2\pi)^{d}}
\frac{e^{-\kappa (p^{2}+m^{2})}}{p^{2}+m^{2}} 
= \int_{\kappa}^{\infty} e^{-\alpha (p^{2}+m^{2})} d\alpha \ee
whose Fourier transform $  C_{\kappa}(x,y)$ is a smooth function
and not a distribution:
\be     C_{\kappa}(x,y) = \int_{\kappa}^{\infty} e^{-\alpha m^{2}- (x-y)^{2}/4\alpha} 
\frac{d\alpha}{\alpha^{D/2}} . \ee
$\alpha^{-D/2} e^{- (x-y)^{2}/4\alpha}$
is the \emph{heat kernel}. Therefore this $\alpha$-representation has also an
interpretation in terms of Brownian motion:
\begin{equation}
C_{\kappa}(x,y) = \int_{\kappa}^{\infty} d\alpha \exp (-m^2 \alpha)\, P(x,y;\alpha)
\end{equation} 
where $P(x,y;\alpha)=(4 \pi \alpha)^{-d/2}
\exp (-{\vert x-y\vert}^2/4\alpha)$ is the Gaussian probability
distribution of a Brownian path going from $x$ to $y$ in time $\alpha$.

Such a regulator $\kappa$ is called
an ultraviolet cutoff, and we have (in the distribution sense)
$\lim_{\kappa \to 0} C_{\kappa}(x,y)= C(x,y)$. Remark that due to the
non zero $m^2$ mass term, the kernel $C_{\kappa}(x,y)$ decays
exponentially at large $\vert x-y\vert$ with rate $m$. 
For some constant $K$ and $d>2$ we have:
\be   \vert C_{\kappa}  (x,y) \vert \le K 
\kappa ^{1-d/2} e^{- m \vert x-y \vert } .  \label{decay}
\ee

It is a standard useful construction to build from the 
Schwinger functions the connected Schwinger functions, given by:

\be  C_{N} (z_{1},...,z_{N}) = \sum_{P_{1}\cup ... \cup P_k = \{1,...,N\}; 
\, P_{i} \cap P_j =0}    (-1)^{k+1}  \prod_{i=1}^{k}  S_{p_i} 
(z_{j_1},...,z_{j_{p_i}}) , \ee
where the sum is performed over all distinct partitions of $\{1,...,N\}$ 
into $k$ 
subsets $P_1,...,P_k$, $P_i$ being made of $p_i$ elements called 
$j_1,...,j_{p_i}$. For instance in the $\phi^4$ theory, where
all odd Schwinger functions vanish due to the unbroken $\phi \to -\phi$ symmetry,
the connected 4-point function is simply:
\bqa C_{4} (z_{1},...,z_{4})  &=&  \label{trunca1}
S_{4} (z_{1},...,z_{4}) - S_{2}(z_{1},z_{2})
S_{2}(z_{3},z_{4}) \\ \no
&& \quad \quad - S_{2}(z_{1},z_{3})S_{2}(z_{2},z_{4}) 
-S_{2}(z_{1},z_{4})S_{2}(z_{2},z_{3})  . \eqa

\subsection{Feynman Graphs}

The full interacting measure may now be defined as the multiplication of
the Gaussian measure $d\mu(\phi)$ by the interaction factor:

\be  d\nu  =  {1 \over Z} e^{-\frac{\lambda}{4!} \int \phi^{4}(x) dx} d\mu(\phi)
\label{mesurenui}\ee
and the Schwinger functions are the normalized moments of this measure:
\be  S_{N} (z_1,...,z_N )  = \int     
\phi(z_1)...\phi(z_N) d\nu (\phi) . \label{schwimom}
\ee

Expanding the exponential as a power series in the coupling constant $\lambda$, 
one obtains a formal expansion for the Schwinger functions:
\be  S_{N} (z_1,...,z_N )  = 
{1 \over Z}  \sum_{n=0}^{\infty} \frac{(-\lambda)^{n}}{n!} \int
\bigl[ \int {\phi^{4}(x) dx \over 4!} \bigr]^{n}  
\phi(z_1)...\phi(z_N)  d\mu (\phi) . \label{schwi}
\ee
It is now possible to perform explicitly the functional integral of
the corresponding polynomial.
The result is at any order $n$ a sum over $(4n+N-1)!!$
Wick contractions schemes $\cW$,  i.e. over all the ways of pairing together 
$4n+N$ fields into $2n+N/2$ pairs. 
The weight or {\it amplitude} of such a scheme $\cW$ is the spatial integral
over $x_1,...,x_n$ of the integrand
$\prod_{\ell \in \cW} C(x_{i_{b(\ell)}},x_i{_{e(\ell)}}) $ 
times the factor 
${1 \over n!}({-\lambda \over 4!})^{n}$. Such amplitudes are functions (in 
fact distributions) of the external positions $z_{1},...,z_N$.
They may diverge either because they are integrals over all of ${\R}^{4}$ (no volume cutoff) 
or because the integrand is typically unbounded due to
the singularities in the propagator $C$ at coinciding points.

\begin{figure}
\centerline{\includegraphics[width=12cm]{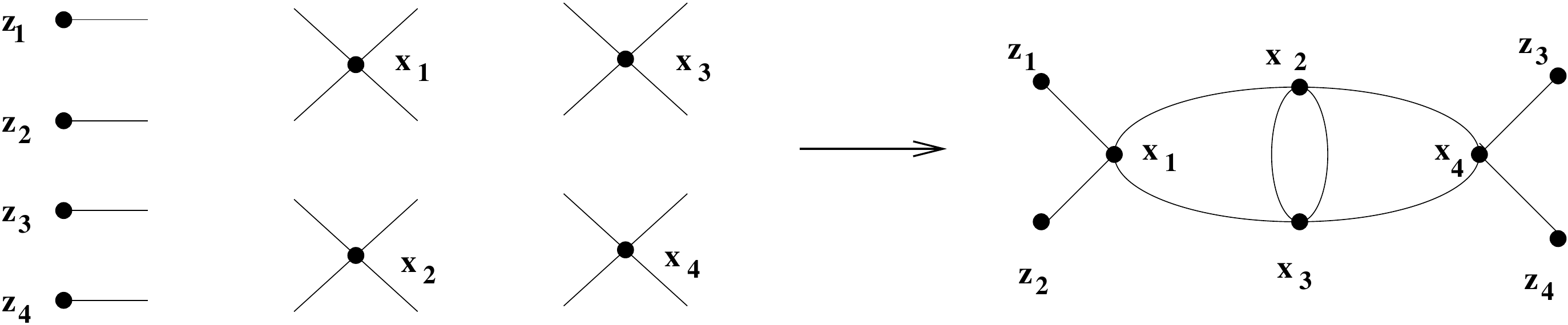}}
\caption{A possible contraction scheme with $n=N=4$.}
\label{graph1}
\end{figure}

Labeling the $n$ dummy integration variables in (\ref{schwi})
as $x_1,...,x_n$, we draw a line $\ell$ for each contraction of two fields.
Each position $x_1,...,x_n$ is then associated to a four-legged vertex
and each external source $z_i$ to a one-legged vertex, as shown in 
Figure \ref{graph1}.

It is convenient to draw these Wick contractions and to regroup all
contractions which give rise to the same drawing or graph.
There are some subtleties about labels. 

\begin{example}
For the normalization at order 1 we have 4 fields, hence 3
Wick contractions, which all give the same graph.
For the 2 point function at order 1 we have 6 fields,
and 15 Wick contractions which fall into 2 categories
with weight 3 and 12.
\end{example}

We have additional observations

\begin{itemize}

\item The great advantage of Feynman graphs is that they form a combinatoric species in the sense of Joyal \cite{BLL} whose logarithm  can be computed as the species of connected graphs. As we already remarked, the computation of
this logarithm is the key physical problem. 

\item However Feynman graphs {\it proliferate}, that is their generating functional
$\sum_n \frac{a_n} {n!} \lambda^n$ has zero radius of convergence in $\lambda$.
At the heart of any constructive strategy \cite{GJ, Riv, LNP, JMP,RivMat,RivMag,GMR}, lies the replacement of the proliferating species of Feynman graphs by a better one \cite{Riv5},
typically  the species of forests. 
The corresponding connected species is the species of trees, which does
not proliferate. Indeed by Cayley's theorem there are only $n^{n-2}$ labeled trees on $n$ vertices. 
This is why constructive expansions converge while ordinary
perturbative expansions dont. Constructive theory ultimately may be considered just as repacking
Feynman graphs in some clever way according to underlying forests \cite{RivWang}. See also
the discussion in subsection \ref{secconstru} and below.

\item The computation factorizes nicely into the connected components
of the graphs. These components may or may not have external arguments. 
In the expansion for the normalized functions the {\it vacuum}  components 
(i.e. those without external arguments) factor out and disappear. Only graphs whose
connected components all contain external arguments remain. 

\item If we further
search for elementary bricks of the expansion, we can consider the
{\it connected} Schwinger functions
like \eqref{trunca1}.
In the expansion of these functions only the graphs
with a single connected component containing all external
arguments survive.

\end{itemize}

\subsection{Feynman Rules}

The ``Feynman rules" summarize how to compute the 
amplitude associated to a Feynman graph with its correct combinatoric factor.

We always use the following notations for a graph $G$:

\begin{itemize}
\item $n(G)$ or simply $n$ 
is the number of internal vertices of $G$, or the 
order of the graph.
\item $l(G)$ or $l$ is the number of internal lines of $G$, i.e. lines 
hooked at both ends to an internal vertex of $G$.
\item $N(G)$ or $N$ is the number of external vertices of $G$; 
it corresponds 
to the order of the Schwinger function one is looking at. 
When $N=0$ the graph 
is a vacuum graph, otherwise it is called an $N$-point graph.
\item $c(G)$ or $c$ is the number of connected components of $G$,
\item $L(G)$ or $L$ is the number of independent loops of G.
\end{itemize}

For a $regular$ $\phi^{4}$ graph, i.e. a graph which has 
no line hooked at both 
ends to external vertices, we have the relations:
\be     l(G) = 2n(G) - N(G)/2 , \label{externu} \ee
\be      L(G)  = l(G) - n(G) + c(G) =  n(G) +1 -N(G)/2 . \label{externu1} \ee
where in the last equality we assume connectedness of $G$, hence $c(G)=1$.

A $subgraph$ $F$ of a graph $G$ is a subset 
of internal lines of $G$, together
with the corresponding attached vertices. 
Lines in the subset defining $F$ are the internal lines 
of $F$, and their number is simply $l(F)$, as before. 
Similarly all the vertices of $G$ hooked to at least one of these 
internal lines of F are called the internal vertices of $F$ and considered 
to be in $F$; their number by definition is $n(F)$. 
Finally a good convention is to call external half-line of $F$
every half-line of $G$ which is not in $F$ but which is hooked to a vertex of 
$F$; it is then the number of such  external half-lines which we call $N(F)$. 
With these conventions one has for $\phi^{4}$ subgraphs the same 
relation (\ref{externu}) as for regular $\phi^{4}$ graphs.

To compute the amplitude associated to a $\phi^{4}$ graph, 
we have to add the contributions of the corresponding contraction schemes. 
This is summarized by the ``Feynman rules":
\begin{itemize}

\item To each line $\ell$ with end vertices at positions $x_\ell$ and $y_\ell$, 
associate a propagator $C(x_\ell,y_\ell)$.
\item To each internal vertex, associate $(-\lambda)/4!$.
\item Count all the contraction schemes giving this diagram. The number 
should be of the form $(4!)^n  n!/S(G)$ 
where $S(G) $ is an integer called the 
symmetry factor of the diagram.  
The $4!$ represents the permutation of the 
fields hooked to an internal vertex.
\item Multiply all these factors, divide by $n!$ and sum over the position 
of all internal vertices.
\end{itemize}

The  formula for the bare amplitude of a graph is therefore, as a distribution 
in $z_1,....z_N$:
\be  A_{G}(z_1,...,z_N) \equiv  \int \prod_{i=1}^{n}
 dx_i  \prod_{\ell \in G} C (x_{i_{b(\ell)}},x_{i_{e(\ell)}}) . \ee
This is the ``direct'' or ``$x$-space'' representation of a Feynman integral.
As stated above, this integral suffers of possible divergences. But 
the corresponding quantity with both volume cutoff $\Lambda$ and ultraviolet cutoff 
$\kappa$, namely
\be  A_{G,\Lambda}^{\kappa}(z_1,...,z_N) \equiv
\int_{\Lambda^{n}} \prod_{i=1}^{n}
dx_i  \prod_{\ell \in G}  C_{\kappa}(x_{i_{b(\ell)}},x_{i_{e(\ell)}}) , \ee
is well defined.
The integrand is indeed bounded and the integration 
domain $\Lambda$ is assumed compact.

The $unnormalized$ Schwinger functions are therefore formally given 
by the sum over all graphs with the right number of external lines of the 
corresponding Feynman amplitudes:

\be ZS_{N} =   \sum_{\phi^{4}{\rm \ graphs \ } G {\rm\ with \ } N(G)=N} 
{(-\lambda)^{n(G)} \over S(G)} A_G
           . \label{series}\ee
$Z$ itself, the normalization, is given by the sum of all vacuum amplitudes:
\be Z =   \sum_{\phi^{4}{\rm \ graphs \ } G {\rm\ with \ } N(G)=0} 
{(-\lambda)^{n(G)} \over S(G)} A_G. \ee

We already remarked that the species of Feynman graphs {\it proliferate} at large orders.
More precisely the total number of $\phi^4$ 
Feynman graphs at order $n$ with $N$ external arguments
is $(4n+N)!!$. Taking into account Stirling's formula and the 
symmetry factor $1/n!$ from the exponential we expect perturbation
theory at large order to behave as $K^n n!$ for some constant $K$. 
Indeed at order $n$ 
the amplitude of a Feynman graph is a 4n-dimensional integral.
It is reasonable to expect that in average it should behave
as $c^n$ for some constant $c$. But this means that one should expect
zero radius of convergence for the series (\ref{series}).
This is not too surprising.
Even the one-dimensional integral 
\be F(g) = \int_{-\infty}^{+\infty}
e^{-x^2/2 - \lambda x^4 /4!} dx 
\ee 
is well-defined only for $\lambda \ge 0$.
We cannot hope infinite dimensional functional
integrals of the same kind to 
behave better than this one dimensional 
integral. In mathematically precise terms,
$F$ is not analytic near $\lambda=0$, but only Borel summable.
Borel summability \cite{Sok}
is therefore the best we can hope for the $\phi^4$ theory,
and we mentioned that it has indeed been established
for the $\phi^4$ theory in dimensions 2 and 3 \cite{EMS,MS}.

From translation invariance, we do not expect $A_{G,\Lambda}^{\kappa}$ to have 
a limit as $\Lambda \to \infty$ if there are vacuum subgraphs in $G$.
But obviously an amplitude factorizes 
as the product of the amplitudes 
of its connected components.

With simple combinatoric verification 
at the level of contraction schemes we can 
factorize the sum over all vacuum graphs in the expansion of 
unnormalized Schwinger functions, hence get for the normalized functions
a formula analog to (\ref{series}):
\be S_N =  \sum_{\scriptstyle \phi^{4}{\rm  \ graphs \ } G {\rm \ with \ }
 N(G)=N
\atop \scriptstyle G {\rm\ without \ any \ vacuum \ subgraph } } 
{(-\lambda)^{n(G)} \over S(G)} A_G . \label{normalfey} \ee
Now in (\ref{normalfey}) 
it is possible to pass to the thermodynamic limit (in the 
sense of formal power series) because using the exponential decrease of the 
propagator, each individual graph has a limit at fixed external arguments. 
There is of course no need to divide by the volume for that because each 
connected component in 
(\ref{normalfey})  is tied to at least one external source, and 
they provide the necessary breaking of translation invariance. 

Finally one can find the perturbative expansions for the 
connected Schwinger functions and 
the vertex functions. As expected, the connected Schwinger functions are given 
by sums over connected amplitudes:
\be  C_N =  \sum_{\phi^{4}{\rm \ connected \ graphs \ } G {\rm \  with \ } 
N(G)=N} {(-\lambda)^{n(G)} \over S(G)} A_{G} 
\ee
and the vertex functions are the sums of the $amputated$ amplitudes for 
proper graphs, also called one-particle-irreducible. 
They are the graphs which remain 
connected even after removal of any given internal line. 
The amputated amplitudes are defined in momentum space by 
omitting the Fourier transform of the propagators 
of the external lines. It is 
therefore convenient to write these amplitudes in the so-called momentum 
representation:
\be  \Gamma_N (z_1,...,z_N)=  \sum_{\phi^{4}{\rm \ proper \ graphs \ } 
G {\rm \  with \ } 
N(G)=N} {(-\lambda)^{n(G)} \over S(G)} A_{G}^T (z_1,...,z_N) , \ee
\be A_{G}^T (z_1,...,z_N) \equiv  { 1 \over  (2\pi)^{dN/2}}
\int dp_1...dp_N  e^{i\sum p_iz_i}  A_G (p_1,...,p_N ) ,  \ee
\be  A_G (p_1,...,p_N ) = \int  \prod_{\ell \ {\rm internal \ line \ of\ }G} 
{d^d p_\ell \over 
p_\ell^2 + m^2}
\prod_{v \in G} \delta ( \sum_\ell \epsilon (v,\ell) \;  p_\ell  )  .
\label{momrep}\ee

Remark in (\ref{momrep}) the $\delta$ functions which 
ensure momentum conservation at 
each internal vertex $v$; the sum inside is over both internal and external 
momenta; each internal line is oriented in an arbitrary way (from $b(\ell)$ to $e(\ell)$) and each 
external line is oriented towards 
the inside of the graph. The incidence 
matrix  $\epsilon(v,\ell)$ captures in a nice way the information
on the internal lines\footnote{Strictly speaking this is true only
for semi-regular graphs, i.e. graphs without tadpoles, i.e. without lines 
which start and end at the same vertex, see \cite{KRTW}.}. It is 1 if the line $\ell$ arrives at $v$, -1 if it starts from $v$ 
and 0 otherwise. Remark also that there is an overall momentum conservation 
rule $\delta(p_1 + ... + p_N)$ hidden in (\ref{momrep}). The drawback of 
the momentum representation lies in the necessity for practical 
computations to eliminate the $\delta$ functions by a ``momentum routing" 
prescription, and there is no canonical choice for that. Although this is 
rarely explicitly explained in the quantum field theory literature, such a choice of a 
momentum routing is equivalent to the choice of a particular spanning tree
of the graph.

\subsection{Scale Analysis and Renormalization}

In order to analyze the ultraviolet or short distance limit according to the renormalization group method \cite{Wil},
we can cut the propagator $C$ into slices $C_i$ so that $C= \sum_{i=0}^{\infty}C_i$. This can be done conveniently within 
the parametric representation, since $\alpha $ in this representation roughly corresponds to
$1/p^2$. So we can define the propagator within a slice 
as 
\be C_{0} =  \int_{1}^{\infty} e^{-m^{2}\alpha -  
{\vert x-y \vert^{2} \over 4\alpha }}  {d\alpha \over \alpha^{d/2}}\ , \ \ 
C_{i} =  \int_{M^{-2i}}^{M^{-2(i-1)}} e^{-m^{2}\alpha -  
{\vert x-y \vert^{2} \over 4\alpha }}  {d\alpha \over \alpha^{d/2}} \ \ {\rm for}\  i\ge 1 .
\label{slicing}
\ee
where $M$ is a fixed number, for instance 10, or 2, or $e$.
We can intuitively imagine $C_i$ as the piece of the field oscillating
with Fourier momenta essentially of size $M^{i}$.
In fact it is easy to prove the bound (for $d>2$)
\be 
\vert C_{i} (x,y) \vert \le K. M^{(d-2)i} 
e^{-  M^{i} \vert x-y\vert }
\label{boundslicing}
\ee
where $K$ is some constant.

Now the full propagator with ultraviolet
cutoff $M^{\rho}$, $\rho$ being a large integer, may be viewed as a sum
of slices:
\be C_{\le \rho} =  \sum_{i=0}^{\rho}  C_{i} . \ee

Then the basic renormalization group step is made of two main operations:

\begin{itemize}
\item A functional integration 
\item The computation of a logarithm
\end{itemize}

Indeed decomposing a covariance in a Gaussian process
corresponds to a decomposition of the field into
independent Gaussian random variables $\phi^i$, each distributed with a
measure $d\mu_i$ of covariance $C_i$. 
Let us introduce
\be \Phi_{i} = \sum_{j=0}^{i} \phi_j .
\ee
This is the ``low-momentum" field for all frequencies
lower than $i$. The RG idea is that starting from
scale $\rho$ and performing $\rho -i$ steps,  
one arrives at an effective action for the remaining field 
$\Phi_{i}$. Then, writing $\Phi_i = \phi_i + \Phi_{i-1}$, one 
splits the field into a ``fluctuation" field
$\phi_i$ and a ``background" field $\Phi_{i-1}$. 
The first step, functional integration, is performed solely
on the fluctuation field, so it computes
\be Z_{i-1} (\Phi_{i-1}) = 
\int d\mu_i (\phi_i) e^{- S_i (\phi_i + \Phi_{i-1}) } .
\label{rengrou1}\ee
Then the second step rewrites this quantity as the exponential 
of an effective action, hence simply computes
\be S_{i-1}  (\Phi_{i-1}) = - \log [ Z_{i-1} (\Phi_{i-1}) ]
\label{rengrou2}\ee
Now $Z_{i-1} =e^{- S_{i-1}}$ and one can iterate!
The flow from the initial bare action $S=S_{\rho}$ for the full field to
an effective renormalized action $S_0$ for the last ``slowly varying" 
component $\phi_0$ of the field is similar to the
flow of a dynamical system. Its evolution
is decomposed into a sequence of discrete 
steps from $S_i$ to $S_{i-1}$.

This renormalization group strategy can be best understood on the system of Feynman graphs
which represent the perturbative expansion of the theory. The first step,
functional integration over fluctuation fields, means that we have to consider 
subgraphs with all their internal lines in higher slices than any of their external lines.
The second step, taking the logarithm, means that we have to consider only
\emph{connected} such subgraphs. We call such connected subgraphs \emph{quasi-local}.
Renormalizability is then a non trivial result that combines
locality and power counting for these quasi-local subgraphs. 

\subsection{Locality, Power Counting}

\begin{figure}
\centerline{\includegraphics[scale=0.4]{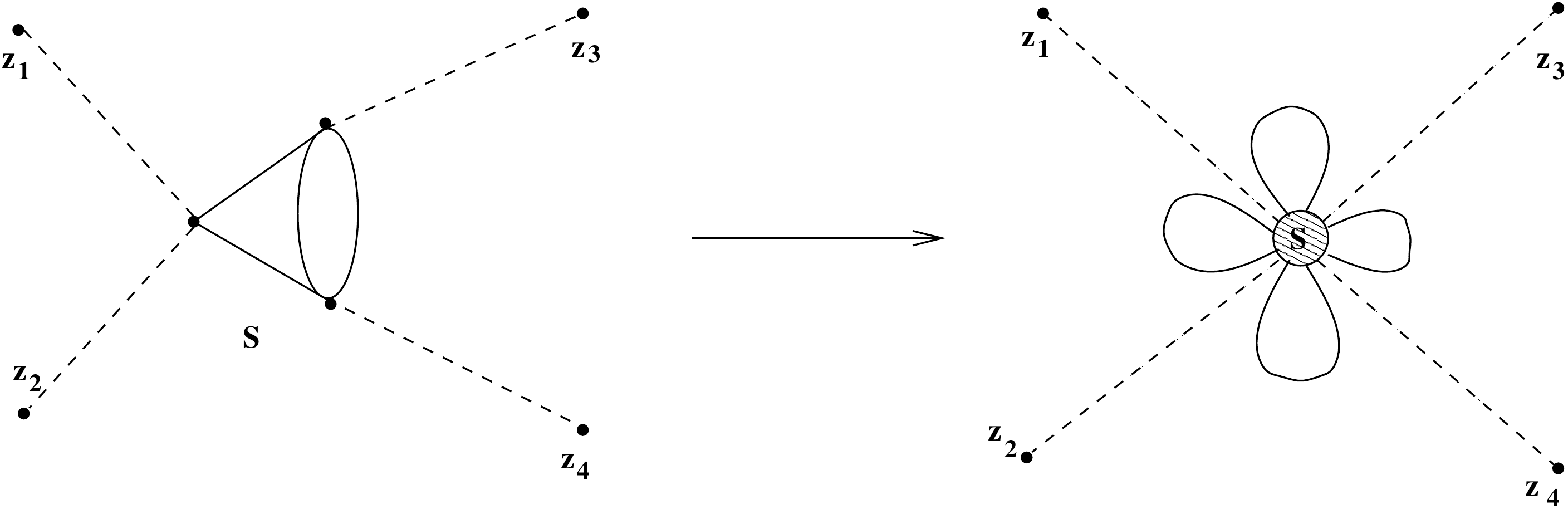}}
\caption{A high energy subgraph {\bf S} seen from lower energies looks quasi-local.}
\label{graph2}
\end{figure}

Locality  simply means that \emph{quasi-local} subgraphs $S$
look \emph{local} when seen through their external lines. Indeed since they are connected and since
their internal lines have scale say $\ge i$,
all the internal vertices are roughly at distance $M^{-i}$.
But the external lines have scales $\le i-1$, which only distinguish details larger than $M^{-(i-1)}$.
Therefore they  cannot distinguish the internal vertices of $S$ one from the other. 
Hence quasi-local subgraphs look like
``fat dots" when seen through their external lines, see Figure \ref{graph2}. 
Obviously this locality principle is completely independent of dimension.

Power counting is a rough estimate which compares the size of a fat dot such as $S$ in Figure \ref{graph2} with $N$ external legs to
the  coupling constant that would be in front of an \emph{exactly local} $\int \phi^N (x) dx$ interaction term if it were in the Lagrangian. To simplify we now assume that the internal scales 
are all equal to $i$, the external scales are 
$O(1)$, and we do not care about constants and so on, but only about the dependence in
$i$ as $i$ gets large. 
We must first save one internal position such as the barycenter of the fat dot or the position of a particular internal vertex to represent the $\int dx$ integration in $\int \phi^N (x) dx$.
Then we must integrate
over the positions of all internal vertices of the subgraph \emph{save that one}. 
This brings about a weight $M^{-di(n-1)}$, because since $S$ is connected 
we can use the decay of the internal lines
to evaluate these $n-1$ integrals. Finally we should not forget the prefactor $M^{(D-2)li}$
coming from (\ref{boundslicing}), for the $l$ internal lines. Multiplying 
these two factors and using relation (\ref{externu})-(\ref{externu1})
we obtain that the "coupling constant" or factor in front of the fat dot
is of order $M^{-di(n-1) +2i(2n-N/2)}= M^{\omega (G)}$, if we define
the superficial degree of  divergence of a $\phi^{4}_{d}$ connected graph as:
\be   \omega (G) = (d-4)n(G) + d - {d-2 \over 2} N(G) .  \ee
So power counting, in contrast with locality, depends on the space-time dimension.

Let us return to the concrete example of Figure \ref{graph2}. 
A 4-point subgraph made of three vertices and four internal lines at a high slice $i$
index. If we suppose the four external dashed lines have much lower index, say of order unity,
the subgraph looks almost local, like a fat dot at this unit scale. We have to save one vertex
integration for the position of the fat dot. Hence
the coupling constant of this fat dot is 
made of two vertex integrations and the four weights of the internal lines (in order not to forget these
internal line factors we kept internal lines apparent
as four tadpoles attached to the fat dot in the right of Figure \ref{graph2}).
In dimension 4 this total weight turns out to be independent of the scale. 

\subsection{Renormalization, Effective Constants}

At lower scales propagators can branch either through the initial bare coupling or through any
such fat dot in all possible ways because of the combinatorial rules of functional integration.
Hence they feel effectively a new coupling which is the sum of the bare 
coupling plus all the fat dot corrections coming from higher scales.
To compute these new couplings only graphs with $\omega (G) \ge 0$,
which are called primitively divergent, really matter
because their weight does not decrease as the gap $i$ increases.

- If $d=2$, we find $  \omega (G) = 2-2n  $, so the only primitively divergent
graphs have $n=1$, and $N=0$ or $N=2$. The only divergence
is due to the ``tadpole'' loop   $\int   {d^2 p \over (p^2 + m^2)}$
which is logarithmically divergent.

- If $d=3$, we find $ \omega (G) = 3- n -N/2  $, so the only primitively divergent
graphs have $n\le 3 $, $N=0$, or $n \le 2$ and $N=2$. 
Such a theory with only a finite number of ``primitively divergent''
subgraphs is called superrenormalizable.

- If $d=4$, $\omega (G) = 4-N$.
Every two point graph is quadratically divergent
and every four point graph is logarithmically divergent.
This is in agreement with the superficial degree of 
these graphs being respectively 2 and 0. 
The couplings that do not decay with $i$ all correspond to terms that 
were already present in the Lagrangian, namely
$\int \phi^4$, $\int \phi^2$ and $\int (\nabla \phi ).(\nabla \phi )$\footnote{Because the graphs 
with $N=2$ are quadratically divergent we must Taylor expand the quasi local fat dots until we get 
convergent effects. Using parity and rotational symmetry, this generates only a
logarithmically divergent $\int (\nabla \phi ).(\nabla \phi )$ term beyond the 
quadratically divergent $\int \phi^2$. Furthermore this term starts only at $n=2$
or two loops, because the first tadpole graph at $N=2$, $n=1$ is \emph{exactly} local.}. 
Hence the structure of the Lagrangian resists under change of scale,
although the values of the coefficients can change.
The theory is called just renormalizable.

- Finally for $d >4$ we have infinitely many primitively divergent graphs
with arbitrarily large number of external legs, and the theory is called
non-renormalizable, because fat dots with $N$ larger than 4 are important and they 
correspond to new couplings generated by the renormalization group  which are not present
in the initial bare Lagrangian.

To summarize:

\begin{itemize}

\item Locality means that quasi-local subgraphs 
look local when seen through their external lines. It holds in any dimension.

\item  Power counting gives the rough size of the new couplings associated to these subgraphs
as a function of their number $N$ of external legs, of their order $n$
and of the dimension of space time $d$.

\item  Renormalizability (in the ultraviolet regime)
holds if the structure of the Lagrangian resists under change of scale,
although the values of the coefficients or coupling constants may change. 
For $\phi^4$ it occurs if $d\le 4$, with $d=4$ the most interesting case.

\end{itemize}

\subsection{The BPHZ Theorem}

The BPHZ theorem is both a brilliant historic
piece of mathematical physics which gives precise mathematical meaning to
the notion of renormalizability, using the mathematics
of formal power series, but it is also ultimately a dead end and a bad way to 
understand and express renormalization. Let us try to explain both statements.

For the massive Euclidean $\phi_4^4$ theory
we could for instance state the following normalization 
conditions on the connected functions in momentum space at zero momenta:
\be C^{4} (0,0,0,0) = -\lambda_{ren} ,        \ee
\be C^{2} (p^{2}=0)  = {1 \over m^{2}_{ren}} , \ee
\be  {d \over dp^{2}} C^{2} \vert _{p^{2}=0}  
= -{a_{ren}\over m^{4}_{ren}} . \ee
Usually one puts $a_{ren}=1$ by rescaling the field $\phi$.

Using the inversion theorem on formal power series for any \emph{fixed ultraviolet cutoff $\kappa$}
it is possible to reexpress any formal power series in $\lambda_{bare}$ with bare propagators
$1/(a_{bare}p^2 +m^2_{bare})$
for any Schwinger functions as a formal power series 
in $\lambda_{ren}$ with renormalized propagators
$1/(a_{ren}p^2 +m^2_{ren})$.
The BPHZ theorem then states that that formal perturbative formal power series  has finite coefficients order by order when the ultraviolet cutoff $\kappa$ is lifted. The first proof by Hepp \cite{Hepp}
relied on the inductive Bogoliubov's recursion scheme \cite{Bop}. Then a completely explicit
expression for the coefficients of the renormalized series was written by
Zimmermann and many followers \cite{Zimm}. The coefficients of that renormalized series
can ne written as sums of renormalized Feynman amplitudes. They are similar
to Feynman integrals but with additional subtractions indexed by 
Zimmermann's forests. Returning to an inductive rather than explicit
scheme, Polchinski remarked that it is possible to also deduce the BPHZ theorem from 
a renormalization group equation and inductive bounds which does not decompose each order of 
perturbation theory into Feynman graphs \cite{Pol}. This method was
clarified and applied by C. Kopper and coworkers, see \cite{kopper}.

The solution of the difficult ``overlapping" divergence problem 
through Bogoliubov's or Polchinski's
recursions and Zimmermann's forests becomes particularly
clear in the parametric representation using Hepp's sectors. A Hepp sector is simply
a complete ordering of
the $\alpha$ parameters for all the lines of the graph. 
In each sector there is a different classification of forests
into packets so that each packet
gives a finite integral \cite{BL,BZ,CR}.

\begin{figure}
\centerline{\includegraphics[width=10cm]{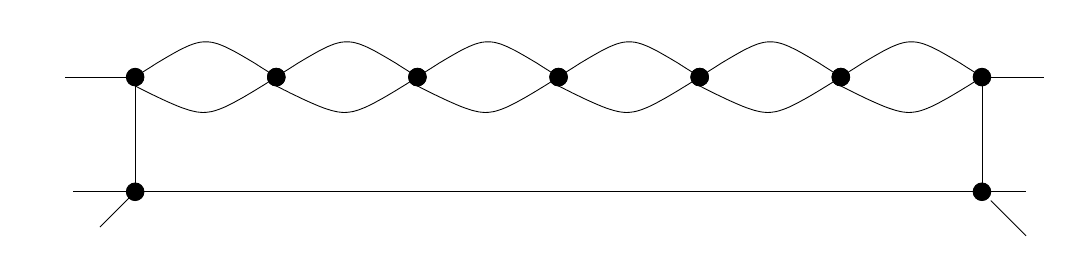}}
\caption{A family of graphs $P_n$ producing a renormalon.}
\label{graph8}
\end{figure}

But from the physical point of view we cannot conceal the fact
that purely perturbative renormalization theory is not very satisfying. 
At least two facts hint at 
a better theory which lies behind:

\begin{itemize}

\item The forest formula seems unnecessarily complicated,
with too many terms. 
For instance in any given 
Hepp sector only one particular packet of forests is 
really necessary to make the renormalized amplitude finite, the
one which corresponds to the quasi-local divergent subgraphs of \emph{that} sector. 
The other packets seem useless, a little bit like ``junk DNA". They are there just because they are necessary for other sectors. 
This does not look optimal.

\item The theory makes renormalized amplitudes finite, but at tremendous cost!
The size of some of these renormalized amplitudes becomes 
unreasonably large as the size of the graph increases. 
This phenomenon is called the ``renormalon problem".
For instance it is easy to check that 
the renormalized amplitude (at 0 external momenta)
of the graphs $P_n$ with 6 external legs and $n+2$ internal 
vertices in Figure \ref{graph8} becomes as large as $c^n n!$ when $n \to \infty$.
Indeed at large $q$ the renormalized
amplitude $A_{G_2}^R$ in Figure \ref{oneloop} grows like $\log \vert
q\vert$. Therefore the chain of $n$ such graphs in Figure \ref{graph8} behaves as
$[\log \vert q\vert]^n$, and the total amplitude of $P_n$
behaves as 
\be  \int 
[\log \vert q \vert]^n  {d^4 q \over [q^2 + m^2 ]^3} \simeq_{n \to\infty} 
c^n n!
\ee
So after renormalization some families of graphs acquire so large values
that they cannot be resummed! Physically this is just as bad as if infinities were still there.
\end{itemize}

\begin{figure}
\centerline{\includegraphics[width=10cm]{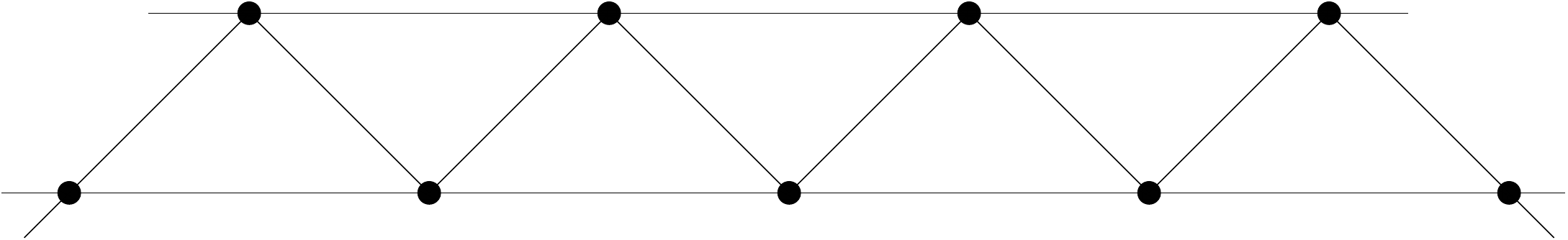}}
\caption{A family of convergent graphs $Q_n$, that do 
not produce any renormalon.}
\label{graph9}
\end{figure}

These two hints are in fact linked. As their name
indicates, renormalons are due to
renormalization. Families of completely convergent graphs such as the graphs
$Q_n$ of Figure \ref{graph9}, are 
bounded by $c^n$, and produce no renormalons. 

Studying more carefully
renormalization in the $\alpha$ parametric representation
one can check that renormalons are solely due
to the forests packets that we compared to ``junk DNA".
Renormalons are due to subtractions that are not necessary to ensure convergence, just like 
the strange $\log \vert q\vert$ growth of $A_{G_0}^R$ at large $q$
is solely due to the counterterm in the region where this counterterm
is not necessary to make the amplitude finite.

We can therefore conclude that subtractions are not organized
in an optimal way by the Bogoliubov recursion. What is wrong 
from a physical point of view in the BPHZ theorem
is to use the size of the graph as the relevant parameter to
organize Bogoliubov's induction. It is rather
the size of the line momenta that should be used to
better organize the renormalization subtractions. 

This leads to the point of view advocated in \cite{Riv}: neither the bare
nor the renormalized series are optimal. Perturbation should be organized
as a power series in an infinite set of effective expansions, which are
related through the RG flow equation. In the end exactly the same contributions are resummed
than in the bare or in the renormalized series, but they are regrouped in a much better
way.

\subsection{The Landau ghost and Asymptotic Freedom}
\label{Landaughost}

In the case of $\phi^4_4$ only the flow of the coupling constants
really matters, because the flow of $m$ and of $a$ for different reasons 
are not very important in the ultraviolet limit:

- the flow of $m$ is governed at leading order by the tadpole.
The bare mass $m^2_i$ corresponding to a finite positive physical mass $m^2_{ren}$
is negative and grows as $\lambda M^{2i}$ with the slice index $i$. But
since $p^2$ in the $i$-th slice
is also of order $M^{2i}$ but without the $\lambda$, as long as the coupling $\lambda$ 
remains small it remains much larger than $m^2_i$. Hence the mass term plays no significant role
in the higher slices. It was remarked in \cite{Riv} that because  there are no
overlapping problem associated to 1PI two point subgraphs, there is in fact no inconvenience
to use the full renormalized $m_{ren}$ all the way from the bare to renormalized scales,
with subtractions on 1PI two point subgraphs independent of their scale.

- the flow of $a$ is also not very important. Indeed it really starts at two loops
because the tadpole is exactly local. So this flow is in fact bounded, and generates no renormalons. 
In fact as again remarked in \cite{Riv} for theories of the $\phi^4_4$
type one might as well use the bare value $a_{bare}$ all the way 
from bare to renormalized scales and perform no second Taylor subtraction on any 
1PI two point subgraphs.

\begin{figure}
\centerline{\includegraphics[width=10cm]{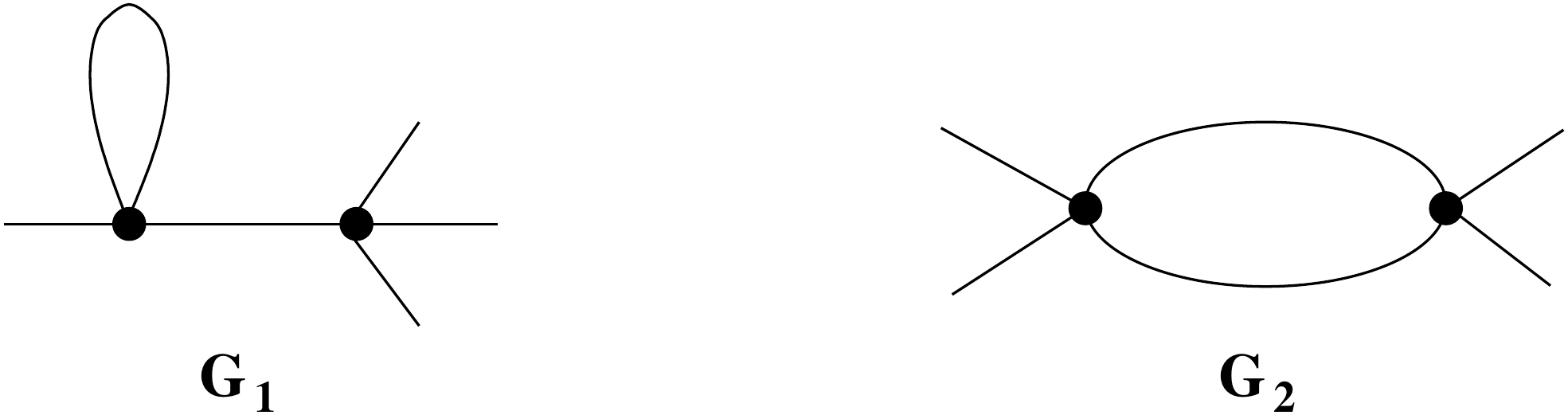}}
\caption{The $\phi^4$ connected graphs with $n=2$, $N=4$.}
\label{oneloop}
\end{figure}

But the physics of $\phi^4_4$ in the ultraviolet limit really depends of the flow
of $\lambda$. By a simple second order
computation there are only 2 connected graphs with $n=2$ and $N=4$
pictured in Figure \ref{oneloop}. They govern at leading order the flow of the coupling constant.

In the commutative $\phi^4_4$ theory the graph $G_1$
does not contribute to the coupling constant flow. This 
can be seen in many ways, for instance after mass renormalization
the graph $G_1$ vanishes exactly because it contains a tadpole
which is not quasi-local but \emph{exactly} local. One can also remark
that the graph is one particle reducible. In ordinary translation-invariant,
hence momentum-conserving theories, one-particle-reducible
quasi-local graphs never contribute significantly to RG flows.
Indeed they become very small when the gap $i$ between internal 
and external scales grows. This is because
by momentum conservation the momentum of any one-particle-reducible line $\ell$ has to be the sum
of a finite set of external momenta on one of its sides.
But a finite sum of small momenta remains small and this clashes directly with the fact that
$\ell$ being internal its momentum should grow as the gap $i$ grows. 
Remark that this is no longer be true in non commutative vulcanized
$\phi^{\star 4}_4$, because that theory is not translation invariant, and that's
why it will ultimately escape the Landau ghost curse.

So in $\phi^4_4$ the flow is intimately linked to the sign
of the graph $G_2$ of Figure \ref{oneloop}. More precisely, we
find that at second order the relation between $\lambda_i$ and $\lambda_{i-1}$
is
\bqa \lambda_{i-1} &\simeq& \lambda_i - \beta \lambda_{i}^2 \label{flow1}
\eqa
(remember the minus sign in the exponential of the action),
where $\beta$ is a constant, namely the asymptotic value of
$\sum_{j, j' / \inf(j,j') =i} \int d^4y C_{j}(x,y)  C_{j'}(x,y) $
when $i \to \infty$. Clearly this constant is positive.
So for the normal stable $\phi_4^4$ theory, the relation
(\ref{flow1}) inverts into
\be \lambda_i \simeq \lambda_{i-1} + \beta  \lambda_{i-1}^2 ,
\ee 
so that fixing the renormalized
coupling seems to lead at finite $i$ to a large, diverging bare coupling,
incompatible with perturbation theory. This is the Landau ghost problem, which affects both the 
$\phi^4_4$ theory and electrodynamics.
Equivalently if one keeps $\lambda_i$ finite as $i$ gets large, $\lambda_0=\lambda_{ren}$
tends to zero and the final effective theory is ``trivial" which means it is a free
theory without interaction, in contradiction with the physical observation
e.g. of a coupling constant of about $1/137$ in electrodynamics.

But in non-Abelian gauge theories an extra
minus sign is created by the algebra of the
Lie brackets. This surprising discovery has deep 
consequences. The flow relation becomes approximately
\be \lambda_i \simeq \lambda_{i-1} - \beta \lambda_i \lambda_{i-1} ,
\ee 
with $\beta > 0 $, or, dividing by $\lambda_i \lambda_{i-1}$,
\be 1/\lambda_i \simeq 1/\lambda_{i-1} + \beta  ,
\ee 
with solution $\lambda_i \simeq {\lambda_0 \over 1 + \lambda_0 \beta i}$. 
A more precise computation to third order in fact
leads to 
\be \lambda_i \simeq {\lambda_0 \over 1 + \lambda_0 (\beta i + 
\gamma \log i + O(1)) }. \ee 
Such a theory is called asymptotically free
(in the ultraviolet limit) because the effective coupling
tends to 0 with the cutoff 
for a finite fixed small renormalized coupling. Physically
the interaction is turned off at small distances. This
theory is in agreement with scattering experiments 
which see a collection of almost free particles
(quarks and gluons) inside the hadrons at very high energy.
This was the main initial argument to adopt
quantum chromodynamics, a non-Abelian gauge
theory with $SU(3)$ gauge group, as the theory of strong interactions \cite{GWil2}. 

Remark that in such asymptotically free theories which form the backbone of today's standard model,
the running coupling constants remain bounded between far ultraviolet ``bare" scales 
and the lower energy scale where
renormalized couplings are measured. Ironically the point of view on early 
renormalization theory as a trick to hide the ultraviolet divergences of QFT
into infinite unobservable bare parameters could not turn out to be more wrong
than in the standard model. Indeed the bare coupling constants
tend to 0 with the ultraviolet cutoff, and what can be farther from infinity than 0?

Recently it has been shown to all orders of perturbation theory that there
should be no Landau ghost but an asymptotically safe fixed
point for the similar RG flow of the non-commutative 
Grosse-Wulkenhaar $\phi^{\star 4}_4$ model  \cite{GW1,GW2,DRbeta,DGMR}. Therefore this model
is a kind of "Ising model" for just renormalizable QFT, that is a simple 
model in which one can presumably fully mathematically
control at last the phenomenon of renormalization in ll its aspects.

\subsection{Grassmann representations of determinants and Pfaffians}

Independent Grassmann variables $\chi_1, ..., \chi_n$ satisfy
complete anticommutation relations
\be \chi_i \chi_j = - \chi_j \chi_i \quad \forall  i,  j
\ee  so that any function of these variables is a polynomial
with highest degree one in each variable. 

The rules of Grassmann integration are defined by linearity and
$$  \int d\chi_i = 0,  \; \;    \;  \;  \int \chi_i d\chi_i = 1 .
$$
plus the rule that all $d\chi$ symbols also anticommute
between themselves and with all $\chi$ variables.

The main important facts are 
\begin{itemize}

\item

Any function of Grassmann variables is a polynomial
with highest degree one in each variable. 

\item

Pfaffians and determinants can be nicely written as Grassmann integrals;

\end{itemize}

The determinant of any $n$ by $n$ matrix can indeed be expressed as
a Grassmann Gaussian integral over $2n$ independent
Grassmann variables which it is convenient to name as
$\bar \psi_1, \ldots , \bar \psi_n$, $\psi_1, \ldots ,  \psi_n$,
although the bars have nothing yet at this stage to do with complex conjugation.
The formula is
\be \det M = \int \prod d\bar \psi_i d\psi_i   e^{-\sum_{ij} \bar\psi_i M_{ij} \psi_j   } .  \label{detgrass}
\ee

Remember  that for ordinary commuting variables and a positive $n$ by $n$ Hermitian matrix $M$ 
\be \frac{1}{\pi^n}\int_{-\infty}^{+\infty} 
\prod_i d\bar \phi_i d\phi_i   e^{-\sum_{ij} \bar\phi_i M_{ij} \phi_j   } = \frac{1}{\det  M} .
\ee

In short Grassmann Gaussian measures are simpler than ordinary Gaussian measures
for two main reasons:

\begin{itemize}

\item
 Grassmann Gaussian measures are associated to any matrix $M$, there is no
positivity requirement for $M$ like for ordinary Gaussian measures. 

\item Their normalization directly computes the determinant of $M$, not the 
inverse (square-root of) the determinant of $M$. This is essential in many areas
where factoring out this determinant is desirable; it explains in particular the success
of Grassmann and supersymmetric functional integrals in the study of {\it disordered systems}.
\end{itemize}

The stubborn reader which remembers the square-root formula in \eqref{gauss1} and would like
to understand the corresponding ``{\it real} version" of \eqref{detgrass} is rewarded by the 
beautiful theory of Pfaffians. Clearly commuting Gaussian real integrals involve symmetric matrices, but Grassmann Gaussian with only $n$ ``real" integrals  must involve $n$ by $n$ antisymmetric matrices.

The Pfaffian $\mathrm{Pf} (A)$ of an \emph{antisymmetric} 
matrix $A$ is defined by
\begin{equation}
\det A=[\mathrm{Pf} (A)]^2 .
\end{equation}
and is known to be polynomila in the coefficients of $A$. This fact is recovered easily
by writing it as
\begin{eqnarray}
\mathrm{Pf} (A) =\int d\chi_1...d\chi_n
e^{-\sum_{i<j}\chi_i A _{ij}\chi_j}
= \int d\chi_1...d\chi_n e^{-\frac{1}{2}\sum_{i,j}\chi_i A _{ij}\chi_j} .
\label{pfaff}
\end{eqnarray}

Indeed we have
\begin{equation}
\det A= \int \prod_i d\bar \psi_i d\psi_i   e^{-\sum_{ij} \bar\psi_i A_{ij} \psi_j   }  .
\end{equation}
Performing the change of variables (which a posteriori justifies the complex notation)
\begin{eqnarray} \label{changepfaff}
\bar\psi_i = \frac{1}{ \sqrt{2}}(\chi_i - i\omega_i), \quad 
\psi_i = \frac{1}{\sqrt{ 2}}(\chi_i + i\omega_i),
\end{eqnarray}
whose Jacobian is $i^{-n}$, the new 
variables $\chi$ and $\omega$ are again independent Grassmann variables.
Now a short computation using $A_{ij}=-A_{ji}$ gives
\bea
\det A&=&  i^{-n}  \int \prod_i d\chi_i d\omega_i   e^{-\sum_{i<j} \chi_i A_{ij} \chi_j 
- \sum_{i<j} \omega_i A_{ij} \omega_j  } \nonumber \\
&=&  \int \prod_i d\chi_i  e^{-\sum_{i<j} \chi_i A_{ij} \chi_j  }\prod_i  d\omega_i   e^{ -\sum_{i<j} \omega_i A_{ij} \omega_j  },
\label{pfaffi}\eea
where we used that $n=2p$ has to be even and that a factor $(-1)^p$ is generated
when changing $ \prod_i d\chi_i d\omega_i $ into $ \prod_i d\chi_i \prod_i d\omega_i $.
Equation (\ref{pfaffi}) shows why $\det A$ is a perfect square and proves (\ref{pfaff}). \qed

A useful Lemma is:

\begin{lemma}\label{quasipfaff}
The determinant of a matrix $D+A$ where $D$ is
diagonal and $A$ antisymmetric has a "quasi-Pfaffian" representation
\be  \det (D+A) = \int \prod_i d\chi_i d \omega_i e^{-\sum_i  \chi_i D_{ii} \omega_i - \sum_{i <j}
\chi_i A_{ij} \chi_j + \sum_{i <j}  \omega_i A_{ij} \omega_j } .
\ee
\end{lemma}
\prf  The proof consists in performing the change of variables 
(\ref{changepfaff}) and canceling carefully the $i$ factors. \qed

There are also {\it normalized} Grassmann Gaussian integrals which may be
expressed formally as 
\be d\mu_M = \frac{\prod d\bar \psi_i d\psi_i   e^{-\sum_{ij} \bar\psi_i M^{-1}_{ij} \psi_j   }}{\int
\prod d\bar \psi_i d\psi_i   e^{-\sum_{ij} \bar\psi_i M^{-1}_{ij} \psi_j   }}.
\ee
and again are characterized by their two point function or covariance
\be \int \bar \psi_i \psi_j d\mu_M =  M_{ij}.
\ee
plus the Grassmann-Wick rule that $n$-point functions are expressed as sum over 
Wick contractions with {\it signs}.

For a much more detailed introduction to the rules of Grassmann calculus in QFT, 
we refer to \cite{Fel}.

\subsection{Trees, forests and the parametric representation}

Classical evolution can be expanded perturbatively into sums indexed by
trees whether in quantum field theory the loops of
Feynman graphs are essential. 

The hidden trees of the classical 
system inside QFT can be revealed only under scale analysis, 
since they do {\it not} correspond to ordinary spanning trees of the graphs, but to
the abstract inclusion relations of short range effects (high energy
quasi local subgraphs) inside larger ones. This point of view
has been progressively formalized over the years
from Bogoliubov to Zimmermann to the most recent formalization by 
D. Kreimer and A. Connes in terms of Hopf algebras.

But {\it ordinary spanning trees} of a connected graph also enter in a fascinating way 
in the computation of its amplitude.
Since the heat kernel is quadratic it is possible to explicitly compute
all spatial integrations in a Feynman amplitude.
One obtains the so-called parametric representation.
The result is expressed in terms of
topological or so-called ``Symanzik'' polynomials \cite{Nak,Sym}.

The amplitude of an 
amputated graph $G$ with external momenta $p$ is, up to a normalization,
in space-time dimension $D$:
\begin{align}
A_G (p) =& \delta(\sum p)\int_0^{\infty} 
\frac{e^{- V_G(p,\alpha)/U_G (\alpha) }}{U_G (\alpha)^{D/2}} 
\prod_l  ( e^{-m^2 \alpha_l} d\alpha_l )\ .\label{symanzik} 
\end{align}
The first and second Symanzik polynomials $U_G$ and $V_G$ are
\begin{subequations}
  \begin{align}
U_G =& \sum_T \prod_{l \not \in T} \alpha_l \ ,\label{symanzik1}\\
V_G =& \sum_{T_2} \prod_{l \not \in T_2} \alpha_l  (\sum_{i \in E(T_2)} p_i)^2 \ , \label{symanzik2}
\end{align}
\end{subequations}
where the first sum is over spanning trees $T$ of $G$
and the second sum  is over two trees $T_2$, i.e. forests separating the graph
in exactly two connected components $E(T_2)$ and $F(T_2)$; the corresponding
Euclidean invariant $ (\sum_{i \in E(T_2)} p_i)^2$ is, by momentum conservation, also
equal to $ (\sum_{i \in F(T_2)} p_i)^2$.

The proof of relations (\ref{symanzik1}-\ref{symanzik2}) is a special case of the 
Tree matrix Theorem, which we now explain following \cite{Abdessel3}

\begin{theorem}[Tree Matrix Theorem]\label{treematrix}
Let $A$ be an $n$ by $n$ matrix such that 
\be \label{sumnulle}
\sum_{i=1}^n A_{ij} = 0 \ \ \forall j \ . 
\ee
Obviously $\det A = 0$. But let $A^{11}$ be the matrix $A$ with line 1 and column 1 deleted.
Then 
\be \det A^{11} = \sum_{T} \prod_{\ell \in T} A_{i_{\ell},j_{\ell}} ,
\ee
where the sum runs over all directed trees on $\{1, ... , n \}$, directed away from root 1. 
\end{theorem}

This theorem has both a positivity and a democracy aspect: all trees
contribute with positive, equal weights to the determinant.

\noindent{\bf Proof of Theorem \ref{treematrix}:}
We use Grassmann variables to write the determinant of a matrix with one line and one raw deleted
as a Grassmann integral with two corresponding sources:
\be
\det A^{11}=
\int ({\rm d}{\Br\psi} {\rm d}\psi)
\ (\psi_1 {\Br \psi}_1)
e^{-{\Br\psi}A\psi}
\ee
The trick is to use (\ref{sumnulle}) to write
\be
{\Br \psi}A\psi=
\sum_{i,j=1}^n
({\Br\psi}_i-{\Br\psi}_j)A_{ij}\psi_j
\ee
and to obtain
\be
\det A^{11} =
\int {\rm d}{\Br\psi} {\rm d}\psi
\ (\psi_1 {\Br \psi}_1)
\exp\lp
-\sum_{i,j=1}^n A_{ij}({\Br\psi}_i-{\Br\psi}_j)\psi_j
\rp
\ee
\be
=
\int {\rm d}{\Br\psi} {\rm d}\psi
\ (\psi_1 {\Br \psi}_1)
\left[
\prod_{i,j=1}^n
\lp 1-A_{ij}({\Br\psi}_i-{\Br\psi}_j)\psi_j \rp
\right]
\ee
by the Pauli exclusion principle. We now expand to get
\be
\det A^{11} =
\sum_{\cG}
\lp
\prod_{\ell=(i,j)\in\cG}(-A_{ij})
\rp
\Om_{\cG}
\ee
where $\cG$ is {\em any} subset of $[n]\times[n]$, and we used the notation
\be
\Om_{\cG}\eqdef
\int {\rm d}{\Br\psi} {\rm d}\psi
\ (\psi_1 {\Br \psi}_1)
\lp
\prod_{(i,j)\in\cG}
\left[ ({\Br\psi}_i-{\Br\psi}_j)\psi_j \right]
\rp
\ee

The tree matrix theorem then follows from the following
\begin{lemma}
$\Om_{\cG}=0$
unless the graph $\cG$
is a tree directed away from 1 in which case
$\Om_{\cG}=1$.
\end{lemma}

\noindent{\bf Proof of the lemma :}
Trivially, if $(i,i)$ belongs to $\cG$, then the integrand of
$\Om_{\cG}$ contains a factor ${\Br\psi}_i-{\Br\psi}_i=0$ and
therefore $\Om_{\cG}$ vanishes. 

But the crucial observation is that if 
there is a loop in $\cG$ then again $\Om_{\cG}=0$.
This is because then the integrand of $\Om_{\cF,\cR}$ contains the factor
\be
{\Br\psi}_{\ta(k)}-{\Br\psi}_{\ta(1)}=
({\Br\psi}_{\ta(k)}-{\Br\psi}_{\ta(k-1)})+\cdots+
({\Br\psi}_{\ta(2)}-{\Br\psi}_{\ta(1)})
\ee
Now, upon inserting this telescoping expansion of the factor
${\Br\psi}_{\ta(k)}-{\Br\psi}_{\ta(1)}$ into the integrand of 
$\Om_{\cF,\cR}$, the latter breaks into a sum of $(k-1)$ products.
For each of these products, there exists an $\al\in\ZZ/k\ZZ$
such that the factor $({\Br\psi}_{\ta(\al)}-{\Br\psi}_{\ta(\al-1)})$
appears {\em twice} : once with the $+$ sign from the telescopic
expansion of $({\Br\psi}_{\ta(k)}-{\Br\psi}_{\ta(1)})$, and once more
with a $+$ (resp. $-$) sign if $(\ta(\al),\ta(\al-1))$
(resp. $(\ta(\al-1),\ta(\al))$) belongs to $\cF$.
Again, the Pauli exclusion principle entails that $\Om_{\cG}=0$.

Now  every connected component of $\cG$ must contain 
1, otherwise there is no way to saturate the $d\psi_1$ integration.

This means that $\cG$ has to be a directed tree on $\{1,... n\}$.
It remains only to see that $\cG$ has to be directed away from 1,
which is not too difficult.
\qed

Relations (\ref{symanzik1}-\ref{symanzik2}) follow rather easily from the tree matrix theorem and the {\it direct
representation of Feynman amplitudes (\ref{paradirect})}. 

In \cite{KRTW} a deeper proof of these relations is given. It relies on the more canonical {\it phase-space
parametric representation}, which we briefly describe now. Let us limit ourselves 
to ``semi-regular" graphs, which have no ``tadpoles" that it no line
starting and ending at the same vertex. These graphs (once their lines have been {\it oriented}
in an arbitrary way) are nicely characterized
by their incidence matrix, which is a regular $l(G)$ by $n(G)$ matrix $\epsilon_{\ell v}$
with 
\bea \epsilon_{\ell v} &=& -1 \;\; {\rm if\;\; line\;\;} \ell\;\;  {\rm exits\;\; vertex}\;\; v  \nonumber \\
 \epsilon_{\ell v} &=& +1 \;\; {\rm if\;\; line\;\;} \ell\;\; {\rm enters\;\; vertex}\;\; v  \nonumber\\
  \epsilon_{\ell v} &=& 0 \;\; {\rm \;\; otherwise\;\;} 
\eea
There are also external momenta $p_f$, $f= 1, \cdots, N$, which we could also also orient through a matrix
$\epsilon_{f v}$.

The momentum parametric representation then writes
\begin{equation}
A^T_G(p_1,...,p_N)= \delta(\sum_{f,v} \epsilon_{fv}  p_f) \int \prod_{\ell =1}^{l(G)}    d \alpha_\ell d^d
k_\ell e^{- \alpha_\ell (k_\ell^2+m^2)} 
\prod_{v=1}^{n(G)-1}  \delta (\epsilon_{fv}  p_f+  \epsilon_{\ell v} k_\ell) .
\label{paramoment}\nonumber
\end{equation}
But there is no canonical way to solve for the delta functions, something known in physics as the procedure
of {\it momentum attribution}. So it is better to rewrite these amplitudes in the
{\it phase-space parametric representation}
\begin{equation}
A^T_G(p_1,...,p_N) = \int  \prod_{\ell =1}^{l(G)} \big[ d \alpha_\ell  e^{-\alpha_\ell m^2} d^d k_\ell  \big]
 \prod_{v=1}^{V-1} d^d x_v
e^{-\alpha_\ell k_\ell^2  + 2 i ( p_f   \epsilon_{vf} x_v   +  k_\ell \epsilon_{v\ell } x_v ) } ,\nonumber
\label{paraphase}
\end{equation}
Integrating over momenta leads  to the
{\it direct space parametric representation}:
\begin{equation}
A^T_G(p_1,...,p_N)=  \int \prod_{\ell =1}^{l(G)}   d \alpha_\ell  
\frac{e^{-\alpha_\ell  m^2}}{\alpha_\ell^{d/2}} 
\prod_{v=1}^{n(G)-1} d^d x_v    e^{2i p_f  \epsilon_{vf} x_v   -  x_v \cdot x_{v'}  \epsilon_{\ell v}  \epsilon_{\ell v'}  / \alpha_\ell  }  .\nonumber
\label{paradirect}
\end{equation}

In \cite{KRTW} it is shown how the representation \eqref{paraphase} together with the quasi-Pfaffian representation of Lemma \ref{quasipfaff}
leads to deletion-contraction relations for the Symanzik polynomials which allow to compute (\ref{symanzik1}-\ref{symanzik2})
from the theory of the universal Tutte polynomial.

\subsection{BKAR Forest Formula} 

Since we want to implement Renormalization Group in a non-perturbative
or constructive way, we need tools to compute connected functions in a non-perturbative
way, with the right scaling properties for the convergence radius of the expansion. For instance
in the just renormalizable case, we need a convergence radius in the coupling constant
which is {\it uniform in the scale index}.

The main such tool is a canonical {\it forest formula} \cite{BK,AR}
which allows to package a perturbative
expansion in terms of trees rather than Feynman graphs. The advantage
was already mentioned several times: the species of trees does not {\it proliferate} \cite{Riv5,RivWang} 
at large orders, in contrast with the species of Feynman graphs.

Consider $n$ points; the set of pairs $P_n$ of such points which has
$n(n-1)/2$ elements $\ell = (i,j)$ for $1\le i < j \le n$, and a smooth function $f$
of $n(n-1)/2$ variables $x_\ell$, $\ell \in \cP_n$. Noting $\partial_\ell$ 
for $\frac{\partial}{\partial x_\ell}$, the standard forest formula is

\be\label{treeformul1}
f(1,\dots ,1)
= \sum_{\cF}  \big[ \prod_{\ell\in \cF}   
\int_0^1 dw_\ell   \big]� \big( [ \prod_{\ell\in \cF} \partial_\ell ] f 
\big)  [ x^\cF_\ell (\{ w_{\ell'}\} ) ]
\ee
where 
\begin{itemize}

\item the sum over $\cF$ is over forests over the $n$ vertices, including the empty one 

\item $x^\cF_\ell (\{ w_{\ell'}\} )$ is the infimum of the $w_{\ell'}$ for $\ell'$
in the unique path from $i$ to $j$ in $\cF$, where $\ell = (i,j)$. If there is no such path,
$x^\cF_\ell (\{ w_{\ell'}\} ) = 0$ by definition.

\item The symmetric $n$ by $n$ matrix $X^\cF (\{w\})$ defined
by $X^\cF_{ii} = 1$ and $X^\cF_{ij} =x^\cF_{ij} (\{ w_{\ell'}\} ) $ 
for $1\le i < j \le n$ is positive.

\end{itemize}

This formula can be viewed as a tool to associate to any 
pair made of a graph $G$ and a spanning forest $F \subset G$ 
a unique rational number or weight $w(G,F)$ between 0 and 1, called 
the relative weight of $T$ in $G$. These weights are barycentric or percentage
factors, ie for any $G$ 
\be  \sum_{F\subset G}  w(G,F) =1  \label{bary}
\ee

The numbers $w(G,F)$ are multiplicative over disjoint unions \footnote{And 
also over vertex joints of graphs, just as in the universality theorem for the Tutte polynomial}.
Hence it is enough to give the formula for $(G,F)$ only when $G$ is {\it connected}
and $F=T$ is a spanning {tree} in it\footnote{It is enough in fact to compute such weights 
for 1-particle irreducible and 1-vertex-irreducible graphs, then multiply them 
in the appropriate way for the general case.}. The definition of these weights is

\begin{definition}
\be  w(G,T) =  \prod_{\ell \in T} \int_0^1 \prod_{\ell \in T} dw_\ell   
\prod_{\ell \not\in T} x^T_{\ell}(\{w\})
\ee
where $x^T_{\ell}(\{w\})$ is again the infimum over the $w_{\ell'}$ parameters 
over the lines $\ell'$ forming the {\it unique} path in $T$ joining the ends of $\ell$.
\end{definition}

Consider the expansion in terms of Feynman amplitudes of a connected quantity $S$.
The most naive way to reorder Feynman perturbation theory according to trees
rather than graphs is to insert for each graph the relation (\ref{bary}) 
\be  S = \sum_{G} A_G = \sum_G \sum_{T\subset G}  w(G,T) \cA_G 
\ee
and exchange the order of the sums over  $S$ and $T$. Hence it writes
\be  S =  \sum_T  \cA_T, \quad  \cA_T = \sum_{G \supset T}  w(G,T) \cA_G .
\ee

This rearranges the Feynman expansion according to trees, but each tree
has the same number of vertices as the initial graph. Hence it reshuffles
the various terms of a {\it given, fixed} order of perturbation theory. Remark
that if the initial graphs have say degree 4 at each vertex, only
trees with degree less than or equal to 4 occur in the rearranged
tree expansion.

For Fermionic theories this is typically sufficient and one has for small enough coupling
\be  \sum_{T}  \vert \cA_T \vert < \infty
\ee
because Fermionic graphs essentially mostly compensate each other
at a fixed order by Pauli's principle; mathematically this is because these
graphs form a determinant and the size of a determinant is much less
than what its permutation expansion usually suggests. This is well known
\cite{Les,FMRT1,AR2}.

But this repacking fails completely for Bosonic theories, because the only compensations there
occur between graphs of different orders. Hence if we perform this naive reshuffling, eg on the 
$\phi^4_0$ theory we would still have
\be  \sum_{T}  \vert \cA_T \vert = \infty .
\ee

Recently a new expansion called the Loop Vertex Expansion has been
found \cite{RivMat} which overcomes this difficulty by exchanging 
the role of vertices and propagators before applying the forest formula. 
It can also be seen as a combination of the forest formula with the so-called
intermediate field representation, which expands into essentially {\it square roots}
of a stable Bosonic interaction. We refer the reader to \cite{RivMat,RivMag,Riv5,RivWang1,RivWang}
but wont review this expansion here, since from now on 
we are mostly going to deal with Fermions.

\subsection{Gram and Hadamard Bounds}

These two bounds on a determinant are often confused! 

The Gram bound applies to a matrix $A = a_{ij}$ whose entries are scalar product.
This means we suppose that there exists some Hilbert space $H$ and $2n$ vectors
$f_i$, $i=1, \cdots , n$, $ g_j$, $j = 1, \cdots , n$ with
\be  a_{ij} = <f_i, g_j>_{H} .
\ee

The Gram bound states
\be    \vert \det A \vert \le \prod_{i=1}^n \Vert f_i \Vert_H  \prod_{j=1}^n  \Vert g_j \Vert_H . \label{gram}
\ee

Of course {\it any} matrix $A$ can always be written of the Gram type, eg with $H = {\mathbb R}^n$,
$f_i = (a_{i,k})$ and $g_j = \delta_{j,k}$, $k=1, \cdots, n$, or conversely. Hence there are two corresponding 
asymmetric Hadamard bounds, one for rows and one for columns:
\be    \vert \det A \vert \le \prod_{i=1}^n \sqrt { \sum_{j=1}^n  a_{ij}^2 } , \label{had1}
\ee
\be    \vert \det A \vert \le \prod_{j=1}^n \sqrt { \sum_{i=1}^n  a_{ij}^2 } \label{had2}
\ee
and also a symmetric Hadamard bound involving the supremum of the matrix elements:
\be    \vert \det A \vert \le   n^{n/2} \bigl( \sup_{i,j} \vert a_{ij} \vert   \bigr)^n . \label{had3}
\ee

Remark that the symmetric Hadamard bound means that a determinant of a large matrix is always 
{\it much smaller} than what its permutation expansion plus naive bounds would suggest, which is
the ``stupid bound"
\be    \vert \det A \vert \le   n ! \bigl( \sup_{i,j} \vert a_{ij}  \vert  \bigr)^n . \label{stupid}
\ee
This difference in constructive theory is essential. Indeed for Fermionic theories with bounded 
propagators and a quartic interaction, the matrix $A$ at $n$-th order of perturbation is a $2n \times 2n$
matrix, with propagators as matrix elements, and there is a $1/n!$ symmetry factor. Hence the bound \eqref{stupid}
would lead to believe that the radius of convergence of the partition function is 0, like in the Bosonic case. But  the Hadamard bound \eqref{had3}
proves
that it is at least finite. Moreover usually it is possible to write the propagators as scalar products in $L^2 ({\mathbb R})^d $ of functions
which also have bounded $L^2$ norms\footnote{This is usually easily done by taking some kind of
``square roots" in momentum space.}. In that case the Gram bound \eqref{gram} shows that the partition function
is in fact an {\it entire} function, as it shows no factorial dependence at all as $n \to \infty$!

\subsection{Single Scale Constructive Theory for a Toy Model}

\label{secconstru}

Consider a just-renormalizable QFT theory. The key problem is to compute connected
quantities with an expansion which converges for a small coupling constant,  with a propagator 
limited to a single renormalization group scale, uniformly in the slice index.
In the simple Fermionic case, this can be done through applying first the BKAR formula
then checking convergence through a Gram bound.

To discuss  the type of Fermionic models met in condensed matter it
is appropriate to consider first a toy Fermionic $d$-dimensional QFT model. It is made of a single an infrared slice with $N$ colors.
Suppose the propagator is diagonal in color space and satisfies the bound
\be \vert C_{j, ab} (x,y) \vert \le  \delta_{ab}  \frac {M^{-dj/2}}{\sqrt N} 
e^{-M^{-j} \vert x-y \vert } .
\ee

We say that the interaction is of the vector type (or Gross-Neveu type) if it is of the form
\be  V = \lambda \int  d^d x 
\bigl(\sum_{a=1}^N\bar\psi_{a}(x)\psi_{a}(x) \bigr)\bigl(\sum_{b=1}^N\bar\psi_{b}(x)\psi_{b}(x)\bigr)
\ee
where $\lambda$ is the  coupling constant.

We claim that

\begin{lemma}
The perturbation theory for the connected functions of this single 
slice model has a radius of convergence in $\lambda$ which is {\it uniform
in $j$ and $N$}.

\end{lemma}

To prove this lemma, expands the partition function $Z(\Lambda)$
through the forest formula, and take the logarithm to obtain a tree
formula for the pressure 
\be p= \lim_{\La \to \infty} {1\over |\La|} \log Z(\Lambda) .
\ee
This is completely straightforward, the only difficulty being
notational. Using the notations of \cite{AR2} 
\bea
p&=&  
\lim_{\La \to \infty} {1\over |\La|}\bigl(
\int d\mu_{C }(\psi, \bar \psi )e^{S_{\Lambda}(\bar\psi_{a}, \psi_{a})}\bigr)
\\
&=&\sum_{n=0}^{\infty}(\la^{n}/N^{n}n!)
\sum_{a_1,\ldots,a_n,b_1,\ldots,b_n=1}^N
\sum_{\cT}\sum_{\Om}\epsilon (\cT,\Om)
\bigl(
\prod_{l\in \cT}\int_{0}^{1} dw_{l} \bigr)
\nonumber\\ \nonumber\
&&\int_{\RR^{nd}} dx_{1}...dx_{n} \de(x_{1}=0)
\prod_{l\in \cT}\bigl(C(x_{i(l)},x_{j(l)})\de_l\bigr)
\times
\det_{remaining} (C_{\alpha\beta})_{\al\in A,\beta \in B}
\label{singexp}
\eea

The sum over the $a_i$'s and $b_i$'s 
are over the colors of the fields and antifields of the vertices obtained by
expanding the interaction and of the form: 
\be
{\bar\psi}_{a_i}(x_i)\psi_{a_i}(x_i)
{\bar\psi}_{b_i}(x_i)\psi_{b_i}(x_i) \label{verint1}
\ee
with $1\le i\le n$.
The sum over $\cT$ is over all trees which connect together
the $n$ vertices at $x_1,\ldots,x_n$.
The sum over $\Om$ is over the compatible ways of realizing the bonds
$l=\{i,j\}\in\cT$ as contractions of a $\psi$ and ${\bar \psi}$ between
the vertices $i$ and $j$ (compatible means that we do not contract twice
the same field or antifield). $\epsilon(\cT,\Om)$ is a
sign which is not important for the bound (see \cite{AR2} for its explicit computation).
For any $l\in\cT$, $i(l)\in\{1,\ldots,n\}$ labels the vertex where the field,
contracted by the procedure $\Om$ concerning the link $l$, was chosen.
Likewise $j(l)$ is the label for the vertex containing the contracted
antifield.
$\de_l$ is 1 if the colors (among $a_1,\ldots,a_n,b_1,\ldots,b_n$)
of the field and antifield contracted by $l$ are the same and else is 0.
Finally the matrix $(C_{\al\beta})_{\al,\beta}$ of the remaining ``loop lines"
is defined in the following manner.

The row indices $\al$ label the $2n$ fields produced by the $n$ vertices,
so that $\al=(i,\si)$ with $1\le i\le n$ and $\si$ takes two values 1 or 2
to indicate whether the field is the second or the fourth factor in \eqref{verint1}
respectively.

The column indices $\beta$ label in the same way the $2n$ antifields,
so that $\beta=(j,\tau)$ with $1\le j\le n$ and $\ta=1$ or 2 according to whether
the antifield is the first or the third factor in \eqref{verint1}  respectively.
The $\al$'s and $\beta$'s are ordered lexicographically. We denote by
$c(i,\si)$ the color of the field labeled by $(i,\si)$ that is
$a_i$, if $\si=1$, and $b_i$ if $\si=2$. We introduce the similar notation
${\bar c}(j,\tau)$ for the color of an antifield.
Now 
\be
C_{(i,\si)(j,\ta)}=w_{ij}^{\cT,BK}({\bf w})C(x_i,x_j)
\de(c(i,\si),{\bar c}(j,\ta)).
\ee
Finally each time a field $(i,\si)$ is contracted by $\Om$
the corresponding row is deleted from the $2n\times 2n$ matrix $(C_{\al\beta})$.
Likewise, for any contracted antifield the corresponding column is erased.
$A$ and $B$ denote respectively the set of remaining rows and the set
of remaining columns. The minor determinant featuring as $\det_{remaining}$ in
\eqref{singexp} is now $\det (b_{\al\beta})_{\al\in A,\beta\in B}$
which is $(n+1)\times(n+1)$.
Indeed for each of the $n-1$ links of $\cT$, a row and a column have been erased.

Suppose we have written
\be
C_j ( x_k, y_m ) <f_{j, k}  , g_{j, m}>_{L^2}
\ee
(this is realized through $f_{j,k} = f_j (x_k, \cdot )$ and 
$g_{j,m} = g_j (\cdot, y_m )$ 
if $\hat f_j (p) . \hat g_j (p) = \hat C_j (p)$).
and that the power counting is conserved by taking square-roots,
namely that the $L^2$ norms of $g_j$ and $f_j$ scale as $ \frac {M^{-dj/4}}{ N^{1/4}} $.

Then applying the Gram inequality
\be
|\det  [C_{j,ab} () ]_{ {\rm remaining }} |   \le
\prod_{ { \rm anti-fields} \;Êk } ||  f_{j,k}   ||
\prod_{{  \rm fields} \;  m} || g_{j,m} ||
\ee
will lead to the proof that the radius of convergence of the pressure is uniform in $j$ and $N$ for this toy model.

Indeed the following lemma shows that the presence of the weakening factors
$w$ does not change the outcome of the Gram bound.

\medskip
\noindent{\bf Lemma}
\medskip
{\it
Let ${\cal A}=(a_{\al\beta})_{\al,\beta}$
be a Gram matrix: $a_{\al\beta}=<f_\al,g_\beta>$ for some inner product
$<.,.>$.
Suppose each of the indices $\al$ and $\beta$ is of the form $(i,\si)$
where the first index $i$, $1\le i\le n$, has the same range as the indices
of the positive matrix $(w_{ij}^{\cF,BK}({\bf w}))_{ij}$, and $\si$
runs through some other index set $\Si$.

Let ${\cal C}=(C_{\al\beta})_{\al,\beta}$ be the matrix with entries
$C_{(i,\si)(j,\ta)}=w_{ij}^{\cF,BK}({\bf w})
.<f_{(i,\si)},g_{(j,\ta)}>
$
and let $(C_{\al\beta})_{\al\in A,\beta\in B}$
be some square matrix extracted from $\cal C$, then for any ${\bf w}$
we have the Gram inequality:
\be 
|\det (C_{\al\beta})_{\al\in A,\beta\in B}|\le
\prod_{\al\in A} ||f_\al||
\prod_{\beta\in B} ||g_\beta ||
\ee
}

\prf  Indeed we can take the symmetric square root $v$ of the 
positive matrix $w^{\cF,BK}$ 
so that $w^{\cF,BK}_{ij} = \sum_{k=1}^n v_{ik}v_{kj} $. 
Let us denote the components of the vectors $f$ and $g$, in an orthonormal
basis for the scalar product $<.,.>$ with $q$ elements, by
$f_{(i,\si)}^m$ and $g_{(j,\ta)}^m$, $1\le m\le q$. (Indeed even if 
the initial Hilbert space is infinite dimensional, the problem is obviously 
restricted to the
finite dimensional subspace generated by the finite set of vectors 
$f$ and $g$).
We then define the tensorized vectors $F_{(i,\si)}$ and $G_{(j,\ta)}$
with components $F_{(i,\si)}^{k m}=v_{i k}f_{(i,\si)}^m$ and
$G_{(j,\ta)}^{k m}=v_{j k}g_{(j,\ta)}^m$ where $1\le k\le n$ and
$1\le m\le q$.
Now considering the tensor scalar product $<.,.>_T$ we have
\be
<F_{(i,\si)},G_{(j,\ta)}>_T=\sum_{k=1}^n\sum_{m=1}^q
v_{ik}v_{jk}f_{(i,\si)}^m g_{(j,\ta)}^m
=b_{(i,\si)(j,\ta)}.
\ee
By Gram's inequality using the $<.,.>_T$ scalar product we get
\be
|\det(C_{\al\beta})_{\al\in A,\beta\in B})|\le
\prod_{\al\in A} {||f_\al||}_T
\prod_{\beta\in B} {||g_\beta ||}_T
\ee
but 
\bea {||F_{(i,\si)}||}_T^2&=&\sum_{k=1}^n\sum_{m=1}^q(F_{(i,\si)}^{km})^2=
\sum_{k=1}^n\sum_{m=1}^q v_{ik}^2(f_{(i,\si)}^m)^2 \nonumber \\
&=&w_{ii}\sum_{m=1}^q(f_{(i,\si)}^m)^2={||f_{(i,\si)}||}^2,
\eea
since $w_{ii}=1$ for any $i$, $1\le i\le n$.
\qed

Let us apply the Gram bound and this last Lemma to bound the determinant in
\eqref{singexp}.

\begin{itemize}

\item
There is a factor $ M^{-dj/2}$ per line, or $ M^{-dj/4}$ per field ie entry of the loop determinant.
This gives a factor $M^{-dj}$ per vertex.

\item
There is a factor $ M^{+dj}$ per vertex spatial integration (minus one)
\end{itemize}

Hence the $\lambda$ radius of convergence is uniform in $j$.

\begin{itemize}

\item
There is a factor $N^{-1/2}$ per line, or $N^{-1/4}$ per field ie entry of the loop determinant.
This gives a factor $N^{-1}$ per vertex

\item
There is a factor $N$ per vertex (plus an extra one)
\end{itemize}

This leads to a bound in
$N. M^{2j}  [c\lambda]^n$ for the $n$-th order of the pressure, hence to
a radius of convergence at least $1/c$. As expected, this is a bound uniform in the slice index $j$.
Hence the $\lambda$ radius of convergence is uniform in $N$.

The last item, namely the factor $N^{n+1}$ for the colors sums is the only one not obvious to prove. Indeed don't know all the graph, but only a spanning tree.
We need to organize the sum over the colors {\it from the leaves to the root of the tree}.
In this way the pay a factor $N$ at each leaf to know the color index which {\it does not go
towards the root}, then prune the leaf and iterate. The last vertex (the root) 
is the only special one as it costs {\it two} $N$ factors.

Let us remark that to treat the corresponding toy model in the Bosonic case
the standard constructive method would be to perform a cluster expansion 
with respect to a lattice of cubes, then a Mayer expansion which further removed the
remaining hardcore constraints with respect to the cubes \cite{Riv}. Both expansions needed to
use the forest formula. This was simplified by the invention of the Loop vertex expansion, 
in which cubes, cluster and Mayer expansions are no longer needed. In addition the
Loop vertex model leads to uniform bounds also for {\it matrix} toy models, a result which cannot be
obtained up to now with other methods \cite{RivMat}.

\section{Interacting Fermions in Two Dimensions}

\subsection{Introduction}

One of the main achievements in renormalization theory has been the extension of
the renormalization group of Wilson (which analyzes long-range behavior
governed by simple scaling around the point singularity
$p=0$ in momentum space) to long-range behavior governed by extended singularities
\cite{BG1,FT1,FT2}. This very natural and general idea is susceptible of many applications 
in various domains, including condensed matter (reviewed here, in which the extended singularity
is the Fermi surface) but also other ones such as
diffusion in Minkowski space (in which the extended singularity is the 
mass shell). In this section we will discuss interacting Fermions models such as those
describing the conduction electrons in a metal.

The key features which differentiate electrons in condensed matter from 
Euclidean field theory, and make the subject in a way 
mathematically richer, is that space-time rotation
invariance is broken, and that particle
density is finite. This finite density of particles 
creates the Fermi sea: particles fill states up to an energy 
level called the Fermi surface. 

The field theory formalism is the best tool to isolate fundamental
issues such as the existence of non-perturbative effects.
In this formalism the usual Hamiltonian point of view
with operators creating electrons or holes
is superseded by the more synthetic point of view of anticommuting 
Fermionic fields with two spin indices  and
arguments in $d+1$ dimensional space-time,. Beware however 
of the QFT-convention to always call dimension the dimension
of {\it space-time}, whether from now on we have to stick to the 
usual condensed matter  convention which is to always call dimension the
dimension of {\it space only}. So one dimensional interacting Fermions
correspond at zero temperature to a two dimensional QFT, two dimensional Fermions
correspond at zero temperature to a three dimensional QFT and so on.

After the discovery of high temperature superconductivity, a key question emerged.
Do interacting Fermions in 2 dimensions (above their low-temperature phase) resemble more three dimensional Fermions, i.e. the Fermi liquid, or one dimensional Fermions, i.e. the Luttinger liquid? The short answer to this controversial question is that it was solved rigorously by mathematical physics and that the answer depends on the shape of the Fermi surface. Interacting Fermions with a round Fermi surface behave more like three dimensional Fermi liquids, whether interacting Fermions with the square Fermi surface of the Hubbard model at half-filling behave more like a one-dimensional Luttinger liquid.

This statement has been now proved in full mathematical rigor, beyond perturbation theory, in the series of works
\cite{DR1,DR2,Hub1,Hub2,Hub3,BGM1,BGM2}.

The existence of usual 2D interacting Fermi liquids was established in \cite{DR1,DR2} 
using the mathematically precise criterion of Salmhofer \cite{Salm2} in the case of a temperature infra-red regulator. Using 
a magnetic field regulator that breaks parity invariance of the Fermi surface
it was also established in the initial sense of a discontinuity at the Fermi surface 
in the series of papers \cite{FKT}.

\subsection{The Models: $J_2$, $J_3$, $H_2$...}

We consider a gas of Fermions in thermal equilibrium at temperature $T$, with coupling 
constant $\lambda$. The free propagator for this model is
\be  \hat C_{ab} (k) = \delta_{ab} \frac{1}{ik_0  - e({\bf k})}
\label{propferm}
\ee
with ${\bf k}$ being the $d$-dimensional momentum, $e({\bf k }) = \epsilon ( {\bf k} ) - \mu $, $\epsilon({\bf k} )$ being the kinetic energy and $\mu$ the chemical potential. 
The $a, b \in  \{\uparrow , \downarrow\}$ index is for spin hence can take two values (remember spin is treated non-relativistically).

At finite temperature, since Fermionic fields have to satisfy antiperiodic boundary
conditions, the component $k_0$ in 
(\ref{propferm}) can take only discrete
values (called the Matsubara frequencies) :
so the integral over $k_0$ is really a discrete sum over $n$. 

These Matsubara frequencies are:
\be k_0 =  \frac{2n+1}{\beta  \hbar} \pi\; , \; \;Ê n \in {\mathbb Z}
\ee
where $\beta =(kT)^{-1}$. 
For any $n$ we have $k_0\neq0$, 
so that the denominator in $C(k)$ can never be 0.
This is why the temperature provides a natural infrared cut-off. 

We can think of $k_0$ as the Fourier dual to an imaginary Euclidean-time continuous variable 
taking values in a circle,
with length proportional to inverse temperature $\beta$.
When $T\to 0^+$, (which means $\beta \to + \infty$), $k_0$ 
becomes a continuous variable, the corresponding discrete
sum becomes an integral, and the corresponding propagator $C_0 (x)$ becomes singular on the Fermi surface defined by $k_0 = 0$ and $e({\bf k}) = 0$. 

This Fermi surface depends on the kinetic energy $\epsilon ( {\bf k} )$ of the model. 

For rotation invariant models, $\epsilon ( {\bf k} )  = {\bf k} ^2 / 2m$ 
where $m$ is some effective or ``dressed" 
electron mass. In this case the energy is invariant
under spatial rotations and the Fermi surface is simply a circle in two dimensions and a
sphere in three dimensions, with radius $\sqrt{2m\mu}$. 
This jellium isotropic propagator is realistic
in the limit of weak electron densities. 
We call this propagator the jellium propagator. We always consider this model, the most natural 
one in the continuum, together with an ultraviolet cutoff (which it is natural to also take rotation invariant)\footnote{ 
The question of whether and how to remove
that ultraviolet cutoff has been discussed extensively in the literature, but we consider it as unphysical
for a non-relativistic model of condensed matter, which is certainly an effective theory at best.}.

Another model considered extensively is the half-filled $2d$ Hubbard model, nicknamed $H_2$. In this model the position variable ${\bf x}$ lives on the lattice ${\mathbb Z}^2$, and $\epsilon ( {\bf k} )  =\cos k_1 +\cos k_2$ so that at $\mu = 0$ the Fermi surface is a square of side size $\sqrt{2\pi}$, joining the points $(±\pi,0),(0,±\pi)$ in the first Brillouin zone.

This propagator is called the Hubbard propagator.

\subsection{Interaction, Locality}

The physical interaction between conduction electrons in a solid 
could be very complicated; the naive Coulomb interaction is in fact subject to heavy screening, the main
effective interaction being due to lattice phonons exchange and other effects.  But we are interested in long-range physics, 
Hence we should use a quasi-local action which decays It is a bit counterintuitive but in fact
perfectly reasonable for a mathematical idealization to use in fact a fully local interaction. This 
should capture all essential mathematical difficulties of the corresponding renormalization group.

But there is a unique exactly local such interaction, namely
\be S_V =\lambda \int_V d^{d+1} \biggl( \sum_{a \in \{\uparrow , \downarrow\} } \bar \psi_a (x) \psi_a (x) 
\biggr)^2 , \label{locact}
\ee
where $V := [-\beta , \beta[ \times V'$ and $V'$ is an auxiliary volume cutoff in two dimensional space, that will be sent to infinity in the thermodynamic limit. Indeed any local polynomial of higher degree is zero since Fermionic fields anticommute. Remark it is of the same form than \eqref{verint}, with spin playing the role of color.

Hence from the mathematical point of view, in contrast with the propagator, the interesting condensed matter interaction is essentially unique. 

The models with jellium propagator and such an 
interaction \eqref{locact}  are respectively 
nicknamed $J_2$ and $J_3$ in dimensions 2 and 3.
The model with Hubbard propagator and interaction \eqref{locact} 
is nicknamed $H_2$.

It is possible to interpolate continuously between 
$H_2$ and $J_2$ by varying the filling factor of the Hubbard model. Lattice models with next-nearest neighbor hopping are also interesting, as they are really the ones used to model the high $T_c$ superconducting phase
in cuprates, but we shall not consider them here for simplicity.

The basic new feature which changes dramatically the power counting
of the theory is that the singularity of the jellium propagator
is of codimension 2 in the $d+1$ dimensional space-time.
Instead of changing with dimension, like in ordinary field theory,
perturbative power counting is now independent of the dimension, and is the
one of a just renormalizable theory. Indeed in a graph with 4 external legs, there are
$n$ vertices, $2n-2$ internal lines and $L=n-1$ independent loops. 
Each independent loop momentum gives rise to two transverse variables,
for instance $k_{0}$ and $|{\bf k} |$ in the jellium case, and to $d-1$ inessential bounded angular variables.
Hence the $2L=2(n-1)$ dimensions of integration for the loop momenta exactly balance
the $2n-2$ singularities of the internal propagators,
as is the case in a just renormalizable theory. 

In one spatial dimension, hence two space-time dimensions,
the Fermi surface reduces to two points, and there is also no proper BCS theory since there is no
continuous symmetry breaking in two dimensions (by the ``Mermin-Wagner theorem''). 
Nevertheless the many Fermion system in one spatial dimension gives
rise to an interesting non-trivial behavior, called the Luttinger liquid \cite{BG}.

The marvel is that although the renormalization group now pinches a non trivial extended singularity,
the locality principle still works. The leading part of the four point quasi-local graphs
is still local, like the interaction of the initial theory.
This is very surprising since quasi-local graphs have internal lines which simply
carry excitations farther from the Fermi surface than their external ones. Momenta close
to the Fermi surface, when moved to a barycenter of the graph, should react through a 
non trivial phase factor. But the miracle is that the only divergent part of the main one-loop
contribution, the ``bubble graph" comes when the combination of the two external legs 
at each end carries approximately zero total momentum. This is because only then can the two inner
bubble propagators both range over the full Fermi sphere. This special configuration, with proper
spin and arrows to ensure no oscillations occur, {\it is really
the Cooper pair}! 

Then these two external legs with approximately zero total momentum can be moved together like a single
low momentum leg in an ordinary Wilsonian renormalization. This is in essence why
renormalization still works in condensed matter and beyond the two point function renormalization 
(Fermi radius renormalization) only changes the value of the coupling constant
for the {\it local} interaction \eqref{locact}. Of course if 
contributions beyond one loop are taken into account, the story becomes more complicated and the renormalization group
flow can in fact involve infinitely many coupling constants \cite{FMRT3}.

A still more surprising case where locality works in a new form, called Moyality, is the case
of non commutative field theory on Moyal space \cite{GW1}. There again the divergent subgraphs are exactly
the only ones that can be renormalized through counterterms of the initial form of the Lagrangian \cite{GMRV}.
This lead us to hope that another still generalized form of locality might hold in quantum gravity.

\subsection{A Brief Review of Rigorous Results}

What did the programs of rigorous mathematical study of interacting Fermi systems accomplish until now? Recall that in dimension 1 there is neither superconductivity nor extended Fermi surface, and Fermion systems have been proved to exhibit Luttinger liquid behavior \cite{BG}. The initial goal of the studies in two or three dimensions was to understand the low temperature phase of these systems, and in particular to build a rigorous constructive BCS theory of superconductivity. The mechanism for the formation of Cooper pairs and the main technical tool to use (namely the corresponding 1/N expansion, where N is the number of sectors which proliferate near the Fermi surface at low temperatures) have been identified \cite{FMRT1}. But the goal of building a completely rigorous BCS theory {\it ab initio} remains elusive because of the technicalities involved with the constructive control of continuous symmetry breaking. So the initial goal was replaced by a more modest one, still important in view of the controversies over the nature of two dimensional Fermi liquids  \cite{And}, namely the rigorous control of what occurs before pair formation. 

As is well known, sufficiently high magnetic field or temperature are the two different ways to break the Cooper pairs and prevent superconductivity. Accordingly two approaches were devised for the construction of ``Fermi liquidsÓ. One is based on the use of non-parity invariant Fermi surfaces to prevent pair formation. These surfaces occur physically when generic magnetic fields are applied to two dimensional Fermi systems. 
In the large series of papers \cite{FKT}, the construction of two dimensional Fermi liquids for a wide class of non-parity invariant Fermi surfaces has been completed in great detail by Feldman, Kn\"orrer and Trubowitz. These papers establish Fermi liquid behavior in the traditional sense of physics textbooks, namely as a jump of the density of states at the Fermi surface at zero temperature, but they do not apply to the simplest Fermi surfaces, such as circles or squares, which are parity invariant.

The other approach is based on Salmhofer's criterion, in which temperature is the cutoff which prevents pair formation.
 The corresponding program studies whether given models satisfy Salmhofer's criterion or not. 
The study of each model has been divided into two main steps of roughly equal difficulty, the control of convergent contributions and the renormalization of the two point functions. In dimension two the corresponding analysis has been completed  for $J_2$, a Fermi liquid in the sense of Salmhofer, and for 
$H_2$ which is not, and is a Luttinger liquid with logarithmic corrections, according to \cite{DR1,DR2,Hub1,Hub2,Hub3}.

Similar results similar have been also obtained for more general convex curves not necessarily rotation invariant such as those of the Hubbard model at low filling, where the Fermi surface becomes more and more circular, including an improved treatment of the four point functions leading to better constants \cite{BGM1,BGM2}. Therefore as the filling factor of the Hubbard model is moved from half-filling to low filling, we conclude that there must be a crossover from Luttinger liquid behavior to Fermi liquid behavior. This sheds light on the controversy \cite{And} over the Luttinger or Fermi nature of two-dimensional many-Fermion systems above their critical temperature.

\subsection{Multiscale Analysis, Angular Sectors}

For any two-dimensional model built until now in the constructive sense, the strategy is the same. It is based on some kind of multiscale expansion, which keeps a large fraction of the theory in unexpanded determinants. The global bound on these determinant (using determinant inequalities such as Gram inequality) is much better than if the determinant was expanded into Feynman graphs which would then be bounded one by one, and the bounds summed. The bound obtained in this way would simply diverge at large order (i.e. not prove any analyticity at all in the coupling constant) simply because there are too many Feynman graphs at large order. But the divergence of a bound does not mean the divergence of the true quantity if the bound is bad. Constructive analysis, which Ókeeps loops unexpandedÓ is the correct way to obtain better bounds, which do prove that the true series in fact does not diverge, i.e. has a finite convergence radius in the coupling constant. This radius however shrink when the temperature goes to 0, and a good constructive analysis should establish the correct shrinking rate, which is logarithmic. This is where multiscale rather than single scale constructive analysis becomes necessary.

The basic idea of the multiscale analysis is to slice the propagator according to the size of its denominator so that the slice with index $j$ corresponds to $\vert i k_0 + e( {\bf k} ) \vert \simeq  M^{-j}$, where $M$ is some fixed constant.

This multiscale analysis is supplemented within each scale by an angular ``sector analysisÓ. The number of sectors should be kept as small as possible, so each sector should be as large as possible in the directions tangent to the Fermi surface in three dimensions, or to the Fermi curve in two dimensions. What limits however the size of these sectors is the curvature of the surface, so that stationary phase method could still relate the spatial decay of a propagator within a sector to its dual size in momentum space. In the case of a circle, the number of sectors at distance $M^{-j}$ of the singularity grows therefore at least like $M^{j/2}$, hence like a power of $T$. However for the half-filled Hubbard model, since the curvature is ``concentrated at the cornersÓ the number of sectors grows only like $| \log T |$. In one dimension there are really only two sectors since the Fermi singularity is made of two points. A logarithm is closer to a constant than to a power; this observation is the main reason for which the half-filled Hubbard model is closer to the one-dimensional Luttinger liquid than to the three dimensional Fermi liquid.

Momentum conservation rules for sectors which meet at a given vertex in general are needed to fix the correct power counting of the subgraphs of the model. In the Hubbard case at half filling, these rules are needed only to fix the correct logarithmic power counting, since the growth of sectors near the singularity is only logarithmic. In both cases the net effect in two dimensions of these conservation rules is to roughly identify two pairs of ÓconservedÓ sectors at any vertex, so that in each slice the model resembles an $N$- component vector model, where $N$ is the number of sectors in the slice.

The multiscale renormalization group analysis of the model then consists essentially in selecting, for any graph, a tree which is a subtree in each of the Óquasi-localÓ connected components of the graph accord- ing to the momentum slicing. These connected components are those for which all internal lines are farther from the Fermi surface than all external lines. The selection of this tree can be performed in a constructive manner, keeping the remaining loop fields in a determinant. The combinatoric difficulty related to the fact that a graph contains many trees is tackled by 
the forest formula.

Once the scale analysis has been performed, a partial
expansion of the loop determinant can detect all the dangerous two and four point functions
which require renormalization. A key point is that this 
expansion can be done without destroying the Gram bound, and the
corresponding sum is not too big (this means its cardinal remains bounded
by $K^n$   (where $K$ is a constant)) because in typical graphs there are not many two and four point
subgraphs.

\subsection{One and Two Particle Irreducible Expansions}

Salmhofer's criterion is stated for the self-energy, i.e. the sum of all one-particle irreducible graphs for the two point function. Its study requires the correct renormalization of these contributions. Since angular sectors in a graph may vary from one propagator to the next in a graph, and since different sectors have different decays in different directions, we are in a delicate situation. In order to prove that renormalization indeed does the good that it is supposed to do, one cannot simply rely on the connectedness of these self- energy graphs, but one must use their particle irreducibility explicitly.
So the proof requires a constructive particle irreducible analysis of the self-energy. The following the-orem summarizes the results of \cite{DR1,DR2}:

\begin{theorem} The radius of convergence of the jellium two-dimensional model perturbative series for any thermodynamic function is at least $c/|\log T |$, where $T$ is the temperature and $c$ some numerical constant. As T and $\lambda$ jointly tend to 0 in this domain, the self-energy and its first two momentum derivatives remain uniformly bounded so that the model is a Fermi liquid in the sense of Salmhofer.
\end{theorem} 

In the case of the jellium model $J_2$, this analysis can be performed at the level of one-particle irreducible graphs \cite{DR2}. The half-filled Hubbard model, however, is more difficult. Although there is no real divergence of the self-energy (the associated counterterm is zero thanks to the particle hole symmetry of the model at half-filling) one really needs a two-particle and one-vertex irreducible constructive analysis to establish the necessary constructive bounds on the self-energy and its derivatives \cite{Hub2}. For parity reasons, the self-energy graphs of the model are in fact not only one-particle irreducible but also two particle and one vertex irreducible, so that this analysis is possible.

This analysis leads to the explicit construction of three line-disjoint paths for every self-energy contribution, in a way compatible with constructive bounds. On top of that analysis, another one which is scale-dependent is performed: after reduction of some maximal subsets provided by the scale analysis, two vertex-disjoint paths are selected in every self-energy contribution. This construction allows to improve the power counting for two point subgraphs, exploiting the particle-hole symmetry of the theory at half-filling, and leads to the desired analyticity result.

Finally an upper bound for the self energy second derivative is combined with a lower bound for the explicit leading self energy Feynman graph \cite{Hub3}. This completes the proof that the Hubbard model violates Salmhofer's criterion, hence is not a Fermi liquid, in contrast with the jellium two dimensional model. More precisely the following theorem summarizes the results of \cite{Hub1,Hub2,Hub3}

\begin{theorem} 
The radius of convergence of the Hubbard model perturbative series at half-filling is at least $c/ \vert \log T \vert $ , where $T$ is the temperature and c some numerical constant. As $T$ and $\lambda$ jointly tend to 0 in this domain, the self-energy of the model does not display the properties of a Fermi liquid in the sense of Salmhofer, since the second derivative is not uniformly bounded.
\end{theorem} 

We would like now to enter into more technical detail, without drowning the reader. Hence we shall
limit ourselves here to the non-perturbative analysis of connected functions in a single
RG scale, which is the core mathematical problem. We compare the various models $J_2$, $J_3$ and $H_2$
to the toy model of section \ref{secconstru}, and explain why sector analysis plus momentum conservation is suited to analyze the
two-dimensional models but fails in three dimensions.

\subsection{2D Jellium Model: Why Sectors Work}

We claim that the $J_2$ model in a slice is roughly similar to the Toy Model, with dimension $d=3$,
provided the momentum slice is divided into angular subpieces called {\it sectors}, which play the role of colors.

The naive estimate on the slice propagator is (using integration by parts)
\be \label{naiveest}
\vert C_{j} (x,y) \vert \le  M^{-j}
e^{-  [M^{-j} \vert x-y \vert ]^{1/2}}
\ee
(using Gevrey cutoffs $f_{j}$ to get fractional exponential decay). The
 prefactor $M^{-j}$ corresponds to the 
volume of integration of the slice, $M^{-2j}$, divided by the slice estimate of
the denominator, $M^{-j}$.
This is much worse than the factor $M^{-3j/2}$ that would be needed.

But the situation improves if we cut the Fermi slice into {\it smaller pieces} (called sectors).
Suppose we divide the $j$-th slice into $M^j$ sectors, each of size roughly $M^{-j}$
in all three directions.

A sector propagator $C^{j,a}$ has now prefactor $M^{-2j}$ corresponding to the 
volume of integration of the sector $M^{-3j}$ divided by the slice estimate of
the denominator, $M^{-j}$. Using integration by parts and  Gevrey cutoffs $f_{ja}$ for fractional power decay
we get without too much effort the bound
\be\vert C_{j, ab} (x,y) \vert \le  \delta_{ab}  M^{-2j}e^{-  [M^{-j} \vert x-y \vert ]^{1/2}} .
\ee

But since $N = M^j$
\be
M^{-2j} = \frac {M^{-3j/2}}{\sqrt N}
\ee
so that this bound is identical to that of the toy model. 

It remains just to explain why the interaction of the model is approximatley of the vector type.
This is because of the momentum conservation rule at every vertex.

 In two dimensions a rhombus (i.e; a closed quadrilateral whose four sides have equal lengths)
is a parallelogram.
Hence an approximate rhombus should be an approximate parallelogram. 

Momentum conservation $\delta (p_1 + p_2 + p_3 + p_4)$  at each vertex follows from translation invariance of $J_2$.
Hence $p_1, p_2, p_3 , p_4$ form a quadrilateral.  For $j$ large 
we have $\vert p_k \vert \simeq  \sqrt{2M \mu}$,
hence the quadrilateral is an approximate rhombus. Hence the four sectors to which $p_1$, $p_2$, $p_3$ and $p_4$
should be roughly equal two by two (parallelogram condition).

It means that the interaction is roughly of the color (or Gross-Neveu) type with respect to these angular sectors:
\be
\bigl( \sum_a   \bar \psi_a  \psi_a \bigr) \bigl( \sum_b  \bar \psi_b  \psi_b \bigr)
\ee

In fact this ``rhombus rule" is not fully correct for almost degenerate rhombuses.
The correct statement is

\begin{lemma}
\label{sectcons1}

Fix $\ m\in {\bf Z}^{3}\ $.
The number of $4$-tuples $
\left\{S_1,\ \cdots\,S_{4}\right\}$
of sectors for which there exist $k_i\in {\mathbb R}^2,\ i=1,\ \cdots,\ 
4\ $ satisfying 
\be
k_i^\prime\in S_i,\ \  | k_i- k_i^\prime|\le {\rm const}\ M^{-j}
\ , \ \ \ i=1,\ \cdots,\ 4\ 
\ee 
and 
\be
 | k_1+\ \cdots\ + k_{4}|\le{\rm const}\left(1+|m|\right)M^{-j} 
\ee
is bounded by 
\be
{\rm const}(1+ | m |)^2 M^{2j}\left\{1+j \right\}.
\ee

\end{lemma}

The $1+j$ factor is special to dimension 2 and is the source of painful technical complications
which were developed by Feldman, Magnen, Trubowitz and myself.

The solution uses in fact $M^{j/2}$ {\it anisotropic} angular sectors, which are
longer in the tangential direction (of length $M^{-j/2}$). The corresponding propagators still
have dual spatial decay because the sectors are still aprroximately flat.

Ultimately the conclusion is unchanged: the radius of convergence of $J_2$ in a slice
is independent of the slice index $j$.

\subsection{Why Sectors Fail in $d=3$}

In 3D sectors and Gram's bound fail by a full power per vertex!

There is indeed no rhombus rule in $d=3$. A closed quadrilateral with equal sides
is not a parallelogram because it can be non-planar; hence it is obtained
by rotating half of a planar parallelogram around the diagonal
by an arbitrary {\it twisting angle}.
Therefore the jellium model interaction is {\it not} of the vector type. More precisely
the analog of Lemma \ref{sectcons1} is, with similar notations

\begin{lemma}

The number of $4$-tuples $
\left\{S_1,\ \cdots\,S_{4}\right\}$
of sectors for which there exist $k_i\in {\mathbb R}^3,\ i=1,\ \cdots,\ 
4\ $ satisfying 
\be
k_i^\prime\in S_i,\ \  | k_i- k_i^\prime|\le {\rm const}\ M^{-j}
\ , \ \ \ i=1,\ \cdots,\ 4\ 
\ee 
and 
\be
 | k_1+\ \cdots\ + k_{4}|\le{\rm const}\left(1+|m|\right)M^{-j} 
\ee
is bounded by 
\be
{\rm const}(1+ | m |)^3 M^{5j}.
\ee
\label{sectcons2}
\end{lemma}
Remark the absence of the log factor that was present in $d=2$.
 But for $d=3$ we find $M^{5j}$ 4-tuples which correspond to the choice of two sectors
($M^{2j} \times M^{2j} $) and of one angular twist $M^j$.
The power counting corresponds to $M^{-3j}$ per sector propagator.
Two propagators pay for one vertex integration ($M^{4j}$) and one
sector choice ($M^{2j}$) but there is {\it nothing to pay for the angular twist}.
Going to anisotropic sectors is possible but there remains still in this case
a $M^j/2$ twist factor. After many years of effort, we concluded that the
sector method and Gram bound apparently cannot be improved to do better.
Hence although $J_3$ is expected to be a Fermi liquid in the sense of Salmhofer, 
new methods have to be developed to treat it constructively \cite{MRHad,DMR,DMR2}.

\subsection{The Hadamard method in $x$-space}

The idea of using Hadamard's inequality in $x$-space to overcome
the constructive power counting problem of Fermions in three dimensions
took Jacques Magnen and myself four years of continuous hard work, with dozens and dozens
of various failed trials, from 1991 to 1995  \cite{MRHad}. Another four or five years took place to fine-tune 
this idea for the multidimensional case with M. Disertori \cite{DMR}. Another ten years have passed to find a momentum-conserving 
version, which should at last allow for the proof of Salmhofer's criterion
in the three dimensional jellium model \cite{DMR2}.

Let us describe the main idea in an informal way.

We remember that the naive estimate \eqref{naiveest} is far from sufficient for correct power counting.
But we also know from perturbative power counting in momentum space that the theory should
be just renormalizable. Therefore the $j$-slice propagator $C^j$ should behave more
as the one of an infrared $\phi^4_4$ theory, hence should be bounded by 
\be \label{typicalest}  K
M^{-2j} e^{-  [M^{-j} \vert x-y \vert ]^{1/2}} .
\ee

In fact this is not totally correct, because it can be shown that the $j$-th slice propagator at almost coinciding points,
hence at $\vert x -y \vert  \simeq 0$ is not bounded by $M^{-2j}$.
Still \eqref{typicalest} is {\it correct at typical distances} for the $J_3$ propagator  in the $j$-th slice. 
We know from eg \eqref{naiveest}  that these typical space-time
distances should be $\vert x -y \vert  \simeq M^j$.

Indeed integrating over angles on the Fermi sphere leads to an $additional $ $1/\vert x - y \vert$ decay, because
\be  \int_{0}^{\pi} \sin \theta d \theta d \phi e^{i \cos \theta \vert x - y \vert }  = \sin  \vert x - y \vert  / \vert x - y \vert  
\ee
Hence for typical distances $\vert x - y \vert  \simeq M^{j}$ the propagator indeed obeys 
the improved estimate
\be  \label{improvest1}
\vert  C_j (x,y)_{\vert x-y \vert \simeq M^j}  \vert   \le K M^{-2j}   e^{-  [M^{-j} \vert x-y \vert ]^{1/2}} .
\ee

The problem is that this bound is wrong at small distances.
However at small distances there is a bonus, namely the corresponding integration volumes are smaller.
Another related problem is that if we use the Gram bound \eqref{gram}
to bound a determinant with all the matrix elements
$C_j (x_p-y_q)$ corresponding to large distances $\vert x_p - y_q \vert$, we still loose
the improvement \eqref{improvest1}, because the $L^2$ norms in \eqref{gram}
will correspond again to propagators at coinciding points!

The solution is to use the Hadamard bounds \eqref{had1}-\eqref{had2} because they conserve the decay of the typical propagators.

Remember however, as seen conveniently in \eqref{had3} that Hadamard bounds consume
the $1/n!$ symmetry factor for $n$ vertices. Hence we can no longer use 
the explicit tree formula and the method of \cite{Les,AR2}.

But we can still use the standard old-fashioned cluster expansion between cubes.
Summarizing:

\begin{itemize}
\item
Just renormalizable power counting is recovered for the main part of the theory
if we use Hadamard's bound rather than Gram's bound.

\item
The factor $n!$ is lost in the Hadamard bound at order $n$; this  forces us to rely on 
the on-canonical tool of cluster expansion between
cubes

\end{itemize}

Still, this solves the constructive problem only for the main part of the propagator, the one at typical distances
$\vert x - y \vert  \simeq M^{j}$ . However at smaller distances there is a bonus, namely the 
volume factors for spatial integration are also smaller.

It turns out that the problem of {\it smaller than typical} distances
can be solved with an auxiliary superrenormalizable decomposition
\be  C^j  = \sum_{k=0}^j  C^{jk}
\ee
of the propagator. Roughly speaking $C^{j,k}$ corresponds to $\vert x-y \vert  \simeq M^k$, 
It means that even the single slice theory in $d \ge 3$
is a non-trivial theory that contains a rather non-trivial 
renormalization, like the one of $\phi^{4}_{3}$. This renormalization
can be analyzed by means of the auxiliary
scales. The solution is in fact more complicated that what we sketch
here, and has to take into account the anisotropy between the space and the imaginary time
variables \cite{MRHad,DMR}.

The use of non-canonical lattices of cubes in this method is the signal that we have probably still not 
found the optimal constructive treatment of $J_3$. Recently we found a better decomposition
that should allow to check Salmhofer's criterion for $J3$ \cite{DMR2}. However in its present stage it will still use
a non-canonical cluster expansion between cubes.
It would be interesting to find a solution such as the loop vertex expansion that solves
the constructive problem in a truly canonical way.

\subsection{2D Hubbard Model}

The Hubbard model lives on the square lattice $\ZZ^2$, so that 
the three dimensional vector $x = (x_0, \x)$ is such that
$ \x = (n_1 , n_2 )\in \ZZ^2$.  
From now on we write $v_1$ and $v_2$ for the two components
of a vector $\vec v$ along the two axis of the lattice. 

At half-filling and finite temperature $T$, the Fourier 
transform of the propagator of the Hubbard model is:
\be
\hat{C}_{ab} (k) = \de_{ab} \frac{1}{ik_0-e({\bf k})},
\quad \quad e({\bf k})= \cos k_1   + \cos k_2 \ ,
\label{prop}
\ee
where $a,b \in \{\uparrow , \downarrow\}$ are the
spin indices. The vector ${\bf k}$ lives on the two-dimensional torus 
$\RR^2/ (2 \pi\ZZ)^2$.
Hence the real space propagator is
\be
C_{ab}(x) =\frac{1}{(2\pi)^2\beta}\; \sum_{k_0} \; 
\int_{-\pi}^{\pi} dk_1 \int_{-\pi}^{\pi}dk_2\; e^{ikx}\;
\hat{C}_{ab}(k) \ .
\label{tfprop}\ee

Recall that $|k_0|\geq \pi/\beta\neq 0$ 
hence the denominator in $C(k)$ again can never be 0 at non zero temperature.
This is why the temperature provides a natural infrared cut-off.
When $T \to 0$ (which means $\beta\to \infty$) $k_0$
becomes a continuous variable, the discrete sum becomes an
integral, and the corresponding propagator $C_{0}(x)$ becomes singular
on the Fermi surface defined by $k_0=0$ and $e({\bf k})=0$. This Fermi surface is a square of side
size $\sqrt{2} \pi$ (in the first Brillouin zone) joining the corners $(\pm \pi , 0), (0,\pm\pi )$.
We call this square the Fermi square, its corners and faces are called
the Fermi faces and corners. Considering the periodic boundary conditions,
there are really four Fermi faces, but only two Fermi corners.

In the following to simplify notations we will write:
\be
\int d^3k \; \equiv \; {1\over \beta} \sum_{k_0} \int d^2k
\quad , \quad 
\int d^3x \; \equiv \; {1\over 2}
\int_{-\beta}^{\beta}dx_0 \sum_{\x\in \ZZ^2} \ . \label{convention}
\ee

In determining the spatial decay we recall that by anti-periodicity
\be
C(x) = f(x_0,\x ) :=
\sum_{m\in \ZZ} (-1)^m \; C_0\lp x_0+{m\over T}, \x \rp  \ .
\label{copie1}\ee
where $C_0$ is the propagator at $T=0$.
Indeed the function $f$ is 
anti-periodic and its Fourier transform is the right one.

The interaction of the Hubbard model is again \eqref{locact}:
\be
S_V = \la  \int_V d^3x\; (\sum_a \bpsi\psi)^2  (x) \label{int} \ ,
\ee
where  $V:= [-\beta,\beta]\times V'$ and $ V'$ is an auxiliary finite volume
cutoff in two dimensional space that will be sent later to infinity.

\subsection{Scale Analysis}

The theory has a natural lattice spatial cutoff. 
To implement the renormalization group analysis, we introduce as usually
a compact support function $u(r)\in{\cal C}_{0}^\infty({\rm R})$
(it is convenient to choose it to be Gevrey of order $\alpha<1$ 
so as to ensure fractional exponential decrease
in the dual space) 
which satisfies:
\be
u(r)= 0 \quad {\rm for} \ |r|> 2 \ ; \ u(r) =1
\quad {\rm for} \   |r|<1  \ \ . \label{gevrey}
\ee
With this function, given a constant $M\ge 2$, we can construct a partition 
of unity
\bqa 1 &=& \sum_{i=0}^{\infty} u_{i} (r)  \ \ \forall r \ne 0\ \ ; 
\nonumber\\ 
u_0 (r)  &=& 1- u(r) \ ;\  u_{i} (r) \ =\ 
u(M^{2(i-1)}r)-u(M^{2i}r) \ {\rm for}\ 
i\ge  1 \ .
\eqa

The propagator is then divided into slices according to this partition
\be
C(k) = \sum_{i=0}^{\infty} C_i(k)
\ee
where
\be
C_i(k) =  C(k)  u_i [ k_0^2+e^2({\bf k}) ] \ .
\ee
(indeed $ k_0^2+e^2({\bf k}) \ge T^{2} > 0$).

In a slice of index $i$ the cutoffs ensure
that the size of $k_0^2 +  e^2({\bf k})$
is roughly $M^{-2i}$. 
More precisely in the slice $i$ we must have
\be M^{-2i} \le k_0^2 +  e^2({\bf k})\le 2M^{2}   M^{-2i} \ .
\label{size}
\ee

\begin{figure}
\centerline{\includegraphics[width=6cm]{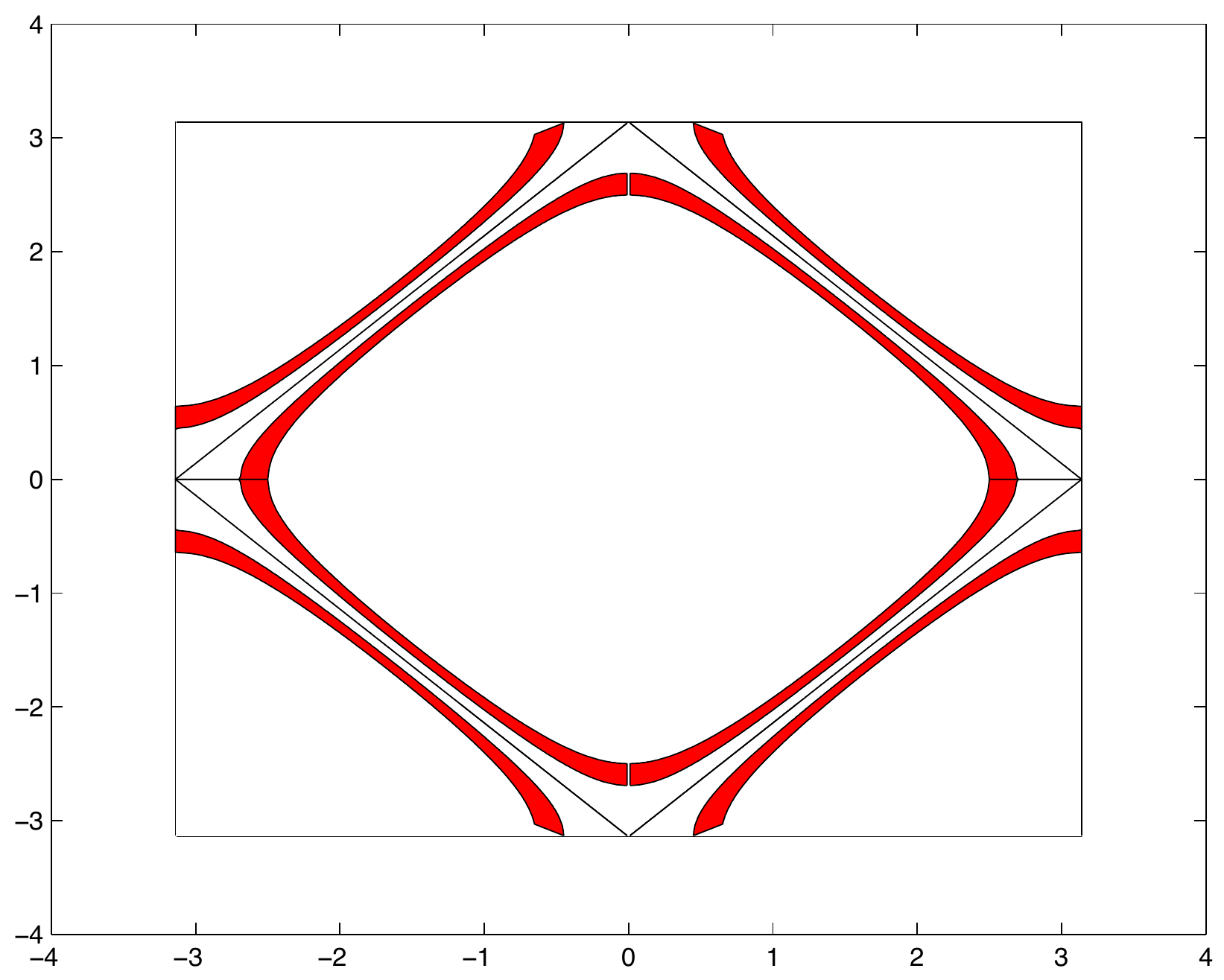}}
\caption{A single slice of the renormalization group}
\label{oslic}
\end{figure}

The corresponding domain is a three dimensional volume whose section
through the $k_0=0$ plane is the shaded region pictured in Figure \ref{oslic}.

Remark that at finite temperature, the propagator $C_i$ vanishes
for $i\ge i_{max}(T)$ where $M^{i_{max}(T)}\simeq 1/T$ (more
precisely  $i_{max}(T) = E( \log {M \sqrt 2 \over \pi T }/\log M)$, 
where $E$ is the integer part), so there is only a finite number of 
steps in the renormalization group analysis. 

Let us state first a simple result,
for a theory whose propagator is only $C_i$, hence corresponding to a generic
step of the renormalization group: 

\begin{theor} The Schwinger functions of the theory with propagator
$C_i$ and interaction (\ref{int}) are analytic in $\lambda$ in a disk
of radius $R_i$ which is at least $c/i$ for a suitable constant $c$:
\be  R_i  \ge c /i \ . 
\ee 
\label{radoneslice}
\end{theor}

The rest of this section is devoted to the definitions and properties 
of $H_2$ sectors, their scaled decay and momentum conservation rules.
As discussed already this result is a first step towards
the rigorous proof \cite{Hub2,Hub3} that $H_2$ is not
a Fermi liquid in the sense of Salmhofer.

\subsection{Sectors}

This section is extracted from \cite{Hub1}.

The "angular" analysis is completely different from the jellium
case. We remark first that in a slice, $k_0^2 + e^2({\bf k})$
is of order $M^{-2i}$, but this does not fix the size of
$e^2({\bf k})$ itself, which can be of order $M^{-2j}$
for some $j \ge i$. In order for sectors defined in momentum space to 
correspond to propagators with dual decay in direct space,
it is essential that their length in the tangential direction is not too big,
otherwise the curvature is too strong
for stationary phase methods to apply. This was discussed first in \cite{FMRT1}.
This leads us 
to study the curve $(\cos k_1 + \cos k_2)^{2} = M^{-2j} $ for arbitrary
$j\ge i$. We can by symmetry restrict ourselves to the region 
$0\le k_1 \le \pi/2$, $k_2 >0$.
It is then easy to compute the curvature radius of that curve, which is

\be  R = {(\sin^2 k_1 + \sin^2 k_2)^{3/2} \over 
| \cos k_1 \sin^2 k_2 + \cos k_2 \sin^2 k_1 |}\ .
\ee
We can also compute the distance $d(k_1) $ 
to the critical curve $ \cos k_1 + \cos k_2 =0$,
and the width $w(k_1)$ of the band 
$M^{-j} \le  |\cos k_1 + \cos k_2 |\le \sqrt{2}M. M^{-j} $.
We can then easily check that 
\be d(k_1) \simeq w(k_1) \simeq {M^{-j} \over M^{-j/2} + k_1 } \ ,
\ee
\be  R(k_1) \simeq   {k_1^3 + M^{-3j/2} \over M^{-j}} \ ,
\ee
where $f\simeq g$ means that on the range $0\le k_1 \le \pi/2$ we have
inequalities $ cf \le g\le df$ for some constants $c$ and $d$.

Defining the anisotropic length 
\be  l(k_1) = \sqrt{ w(k_1)R(k_1) } \simeq M^{-j/2} + k_1 \ ,
\ee
the condition in \cite{FMRT1} is that the sector length should
not be bigger than that anisotropic length. This leads to the idea that
$k_1$ or an equivalent quantity should be sliced according to a geometric
progression from 1 to $M^{-j/2}$ to form the angular sectors in this model.

For symmetry reasons
it is convenient to introduce a new orthogonal but not normal 
basis in momentum space $(e_+, e_-)$,
defined by $e_+ = (1/2)(\pi, \pi)$ and $e_- = (1/2)(-\pi, \pi)$.
Indeed if we call $(k_+, k_{-})$ the coordinates
of a momentum $k$ in this basis, the Fermi surface is given by
the simple equations $k_+ = \pm 1$ or  $k_- = \pm 1$.
This immediately follows from the identity
\be \cos k_1 + \cos k_2  = 2 \cos (\pi  k_{+}/2 )
\cos (\pi  k_{-}/2 ) \ .
\ee
(Note however that the periodic b.c. are more complicated in that new basis).
Instead of slicing $e({\bf k} ) $ and $k_1$, it is then more symmetric to
slice directly $\cos (\pi  k_{+}/2 )$ and $\cos (\pi  k_{-}/2 )$.

Guided by these considerations we introduce the partition of unity
\be
1 = \sum_{s=0}^{i} v_{s}(r) \ ; 
\begin{cases} v_{0}(r) = 1- u(M^{2}r) \\
v_{s}= u_{s+1}\ & {\rm for}  1\le s\le i-1 \\
v_{i}(r) = u(M^{2i}r) 
\end{cases}
\ee
and define
\be
C_{i}(k) = \sum_{\sigma = (s_{+}, s_{-})} C_{i,\sigma}(k)
\ee
where
\be
C_{i,\sigma}(k)=  C_i(k) v_{s_{+}} [ \cos^{2} (\pi k_{+} /2 ) ]\;
v_{s_{-}} [ \cos^{2}\pi k_{-}/2) ] \ .
\ee
We remark that using (\ref{size}) in order for $C_{i,\sigma}$ not to be 0, 
we need to have $s_{+} + s_{-} \ge i-2$. We define the ``depth''
$l(\sigma)$ of a sector to be $l = s_{+} + s_{-} - i + 2$.

To get a better intuitive picture of the sectors,
we remark that they can be classified into different categories:

\begin{itemize}

\item  the sectors (0,i) and (i,0) are called the middle-face sectors

\item the sectors (s,i) and (i,s) with $0 <s<i$ are called the face sectors

\item the sector (i,i) is called the corner sector

\item the sectors (s,s) with $ (i-2)/2 \le s <i$ are called the diagonal sectors

\item the others are the general sectors

\end{itemize}

Finally the general or diagonal
sectors of depth 0 for which $s_{+} + s_{-} = i-2$ 
are called border sectors.

If we consider the projection onto the $(k_{+}, k_{-})$ plane,
taking into account the periodic b.c. of the Brillouin zone,
the general and diagonal sectors have 8 connected components, the 
face sectors have 4 connected components, the middle
face sectors and the corner sector have 2 connected components. 
In the three dimensional space-time, 
if we neglect the discretization of
the Matsubara frequencies, these numbers would double 
except for the border sectors.

\subsection{Scaled decay}

\begin{lemma} Using Gevrey cutoffs of degree $\alpha <1$, 
the propagator $C_{i,\si}$ obeys the scaled decay
\be | C_{i,\si} | \le  c. M^{-i-l} e^{-[d_{i,\si } (x,y)]^{\alpha}}
\label{decay1}
\ee
where
\be  d_{i,\si } (x,y) = \{ M^{-i}|x_0 -y_0| +  
M^{-s_{+} }|x_{+} -y_{+}| +
M^{-s_{-}}|x_{-} -y_{-}| \}  \ .
\ee
\label{Gevreydecay2}
\end{lemma}

\noindent{\it Proof} This is essentially Fourier analysis 
and integration by parts.
If $x = (n_1, n_2)\in \ZZ^{2}$, we define $(x_{+}, x_{-})= (\pi/2)(n_1+n_2,
n_2 -n_1)$. The vector $(x_{+}, x_{-})$ then belongs to $(\pi/2)\ZZ^{2}$
but with the additional condition that $x_{+}$ and $x_{-}$ have the same 
parity. 

Defining, for $X \in [(\pi/2)\ZZ\;]^{2}$
\bqa  D_{i,\sigma}( X) = && (1/2) {1 \over 8 \beta} \sum_{k_0}
\int_{-2}^{+2} dk_{+}\int_{-2}^{+2} dk_{-}
e^{i (k_{0}x_{0}+ k_{+}x_{+} + k_{-}x_{-})}  
\nonumber \\
&& {u_i [ k_0^2+  4\cos^{2}(\pi k_{+}/2) \cos^{2} (\pi k_{-}/2) ] 
\over ik_{0} - 2\cos(\pi k_{+}/2) \cos (\pi k_{-}/2) } 
\nonumber \\
&& v_{s_{+}} [ \cos^{2} (\pi k_{+} /2 ) ]\;
v_{s_{-}} [ \cos^{2}(\pi k_{-}/2) ]
\label{doubl}
\eqa
we note that $C_{i,\sigma}(X)= D_{i,\sigma}( X)$ for $X$ satisfying
the parity condition.

(Remember the Jacobian ${\pi ^{2} \over 2}$ from $dk_1 dk_2$ to 
$dk_+ dk_-$, and the initial domain of integration that is doubled.)

The volume of integration trivially gives a factor $M^{-i}$ for the 
$k_{0}$ sum and factors $M^{-s_{+}}$ and $M^{-s_{-}}$ for the
$k_{+}$ and $k_{-}$ integration (see (\ref{supp}) below). The integrand
is trivially bounded by $M^{i}$ on the integration domain, and this explains
the prefactor $cM^{-i-l}$ in (\ref{decay1}).

We then apply standard integration by parts techniques to formulate
the decay. From e.g. Lemma 10 in \cite{DR1} we know that to obtain
the scaled decay of Lemma \ref{Gevreydecay2} we have only to
check the usual derivative bounds in Fourier space:

\bqa \Vert {\partial^{n_0} \over \partial k_0^{n_0} }
{\partial^{n_+} \over \partial k_+^{n_+} }
{\partial^{n_-} \over \partial k_0^{n_-} } \hat D_{i,\sigma} \Vert
\le A.B^{n} M^{i n_0} M^{s_{+}n_+} M^{s_{-}n_-} 
(n !)^{1/\alpha} \label{foubou}
\eqa
where $n=n_{0}+n_{+}+n_{-}$, and the derivative 
${\partial \over \partial k_0}$ really means the 
natural finite difference operator $(1/2\pi T) (f(k_0 +2\pi T)-f(k_0))$ 
acting on the discrete Matsubara frequencies. The norm is the ordinary sup
norm.
 
But from (\ref{doubl}), 
\bqa\hat D_{i,\sigma} (k)&=& {1 \over 16\beta}  
{u_i [ k_0^2+  4\cos^{2}(\pi k_{+}/2) \cos^{2} (\pi k_{-}/2) ] 
\over ik_{0} - 2\cos(\pi k_{+}/2) \cos (\pi k_{-}/2) } 
\nonumber\\
&&v_{s_{+}} [ \cos^{2} (\pi k_{+} /2 ) ]\;
v_{s_{-}} [ \cos^{2}(\pi k_{-}/2) ]
\eqa
and the derivatives are bounded easily using the standard
rules for derivation, product and composition of Gevrey functions,
or by hand, using the support properties of the 
$v_{s_{+}}$ and $v_{s_{-}}$ fonctions. For instance a derivative
${\partial \over \partial k_+}$ can act on the 
$v_{s_{+}} [ \cos^{2} (\pi k_{+} /2 ) ]$ factor, in which case it 
is easily directly bounded by $c M^{s_{+}}$ for some constant $c$. 
When it acts on
$u_i [ k_0^2+  4\cos^{2}(\pi k_{+}/2) \cos^{2} (\pi k_{-}/2) ] $
it is easily bounded by  $c.M^{2i-s_+-2s_-}$ 
hence by $c. M^{s_{+}}$, using the relation $s_{+}+ s_{-}\ge i-2$.
When it acts on the denominator
$[ik_{0} - 2\cos(\pi k_{+}/2) \cos (\pi k_{-}/2) ]^{-1} $.
it is bounded by $c.M^{i-s_-}$, hence again 
by $c. M^{s_{+}}$, using the relation $s_{+}+ s_{-}\ge i-2$.
Finally when it acts on a  $\cos (\pi k_{+} /2 ) $ created
by previous derivations, it costs directly $c. M^{s_{+}}$.
The factorial factor $(n !)^{1/\alpha}$ in (\ref{foubou})
comes naturally from deriving the cutoffs,
which are Gevrey functions of order $\alpha$; deriving other factors
give smaller factorials (with power 1 instead of $1/\alpha$).

Finally a last remark: to obtain the Lemma for the last slice, 
$i=i_{max}(T)$, one has to take into account the fact 
that $x_{0}$ lies in a compact circle, so that there is really no 
long-distance decay to prove.

\subsection{Support Properties}

If $C_{i,\si } (k) \ne 0$, the momentum $k$ must obey
the following bounds:

\be |k_0| \le \sqrt 2 M  M ^{-i}
\ee
\be \begin{cases} M^{-1}  \le  |\cos (\pi k_{\pm} /2)| \le 1
\ & {\rm for} \;\; s_{\pm} = 0   \ ,\\
M^{-s_{\pm}-1}  \le  |\cos (\pi k_\pm /2)| \le \sqrt 2 M ^{-s_{\pm}}
\ & {\rm for} \;\;  1\le s_{\pm}\le i-1  \ ,\\
|\cos (\pi k_\pm /2)| \le \sqrt 2 M ^{-i}
\ &  {\rm for} \;\;  s_{\pm} = i \ . \end{cases}
\ee
In the support of our slice in the first
Brillouin zone we have $ |k_+ | < 2$ and 
$|k_-| < 2$ (this is not essential but the inequalities 
are strict because $i\ge1$). 
It is convenient to associate to any such component
$k_{\pm}$ a kind of ``fractional part'' called $q_{\pm}$ defined by
$q_{\pm}= k_{\pm}-1$ if $k_{\pm}\ge 0$ and  
$q_{\pm}= k_{\pm}+1$ if $k_{\pm}<0$, 
so that $0\le |q_{\pm}| \le 1 $. 
Then the bounds translate into
\be \begin{cases}  2/ \pi M   \le | q_\pm  | \le 1
\ &  {\rm for} \;\;   s_{\pm} = 0   \ ,\\
2 M^{-s_{\pm}} / \pi M  \le | q_\pm  | \le \sqrt 2
M ^{-s_{\pm }} 
\ &  {\rm for} \;\;  1\le s_{\pm }\le i-1  \ ,\\
| q_\pm   |\le \sqrt 2  M ^{-i} 
\ &  {\rm for} \;\;   s_{\pm } = i  \ . \end{cases}
\label{supp}
\ee

\subsection{Momentum conservation rules at a vertex}

Let us consider that the four momenta $k_1$, $k_2$, $k_3$, $k_4$,
arriving at a given vertex $v$
belong to the support of the four sectors 
$\si_1$, $\si_2$, $\si_3$, $\si_4$, in slices $i_1$, $i_2$, $i_3$, $i_4$.
In Fourier space the vertex (\ref{int}) implies constraints
on the momenta. Each spatial component of the sum of the four momenta 
must be an integer multiple of $2\pi$ in the initial basis, 
and the sum of the four Matsubara frequencies must also be zero.
 
In our tilted basis $(e_{+},e_{-})$, this translates into the conditions:

\be k_{1,0} + k_{2,0} + k_{3,0} + k_{4,0} = 0 \ ,
\ee
\be k_{1,+} + k_{2,+} + k_{3,+} + k_{4,+} = 2n_{+}  \ ,\label{consplus}
\ee
\be k_{1,-} + k_{2,-} + k_{3,-} + k_{4,-} = 2n_{-}  \ ,\label{consminus}
\ee
where $n_{+}$ and $n_{-}$ must have identical parity.

We want to rewrite the two last equations in terms of the 
fractional parts $q_{1}$, $q_{2}$, $q_{3}$ and $q_{4}$. 

Since an even sum of integers which are $\pm 1$ is even, we find
that (\ref{consplus}) and (\ref{consminus}) imply  
\be q_{1,+} + q_{2,+} + q_{3,+} + q_{4,+} = 2m_{+}  \ ,\label{coplus}
\ee
\be q_{1,-} + q_{2,-} + q_{3,-} + q_{4,-} = 2m_{-}  \ ,\label{cominus}
\ee
with $m_{+}$ and $m_{-}$ integers. Let us prove now that except
in very special cases, these integers must be 0. Since $|q_{j,\pm}|\le 1$,
$|m_{\pm}|\le 2$. But  $|q_{j,\pm}| = 1$ is possible only for $s_{j,\pm}=0$.
Therefore $|m_{\pm}| = 2$ implies $s_{j,\pm}=0 \ \forall j$.
Now suppose e.g. $|m_{+}|=1$. 
Then  $s_{j,+}$ is 0 for at least two values of 
$j$. Indeed for $s_{j,\pm}\ne 0$ we have $|q_{j,\pm}|\le \sqrt 2 M^{-1}$,
and assuming $3 \sqrt 2 M^{-1} < 1$, equation (\ref{coplus}) could not hold.

We have therefore proved

\begin{lemma}
$m_{+}=0$ unless $s_{j,+}$ is 0 for at least two values of 
$j$, and $m_{-}=0$ unless $s_{j,-}$ is 0 for at least two values of 
$j$.
\end{lemma}

Let us analyze in more detail equations (\ref{coplus}) and 
(\ref{cominus}) for $|m_{+}|=|m_{-}|=0$. Consider e.g.
(\ref{coplus}). By a relabeling
we can assume without loss of generality that
$s_{1,+} \le s_{2,+} \le s_{3,+}\le s_{4,+}$
Then either $s_{1,+}= i_1 $ or $s_{1,+} < i_1$, in which case
combining equations (\ref{coplus}) and (\ref{supp}) we
must have:
\be 3\sqrt 2 M^{-s_{2,+}} \ge 2 M^{-s_{1,+}} /\pi M \ ,
\ee
which means
\be s_{2,+} \le s_{1,+} + 1 + {\log (3\pi/\sqrt 2) \over \log M} \ .
\ee
This implies
\be | s_{2,+} - s_{1,+}| \le 1 
\ee
if $M > 3 \pi /\sqrt 2 $, which we assume from now on.

The conclusion is:

\begin{lemma}
If $m_{\pm}=0$, either the smallest index $s_{1,\pm}$ 
coincides with its scale $i_1 $, or the two smallest
indices among $s_{j,\pm}$ differ by at most one unit.
\end{lemma}

Now we can summarize the content of both Lemmas in a slightly weaker 
but simpler lemma:

\begin{lemma}
\noindent

\noindent{\bf A) (single slice case)}

\medskip
The two smallest
indices among $s_{j,+}$ for $j=1,2,3,4$ differ by at most one unit,
and the two smallest
indices among $s_{j,-}$ for $j=1,2,3,4$ differ by at most one unit.

\medskip
\noindent
{\bf B) Multislice case}

The two smallest
indices among $s_{j,+}$ for $j=1,2,3,4$ differ by at most one unit
or the smallest one, say $s_{1, +}$ must coincide with its scale $i_1$, 
which must then be strictly smaller than the three other scales $i_2$, $i_{3}$
and $i_{4}$. Exactly the same statement holds independently 
for the minus direction.
\label{consmom}
\end{lemma}

This lemma allows to check that again the single scale analysis works and
leads to a radius of convergence independent of the slice \cite{Hub1}.

\subsection{Multiscale Analysis}

The half-filling point is very convenient since the particle-hole exact symmetry at this point
ensures that there is no flow for the Fermi surface itself.

Using the sector decomposition and the momentum conservation we find
that power counting of the 2D Hubbard model is essentially similar to one dimensional
case with logarithmic corrections \cite{Hub2}. Expanding the two point function to second order
we can even find {\it lower} bounds which prove that this model is {\it not} a Fermi liquid
in the sense of Salmhofer \cite{Hub3}.

\subsection{Acknowledgments}
I thank J. Magnen, M. Disertori, M. Smerlak and L. Gouba for 
contributing various aspects of this work.

\end{document}